\documentclass[final,1p,times]{elsarticle}
\usepackage{amssymb,amsmath}
\usepackage{subeqn}
\begin{document}
%%%%%%%%%%%%%%%%%%%%%%%%%%%%%%%%%%%%%%%%%%%%%%%%%%%%%%%%%%%%%%%%%%%%%%%%%%%%%%%%%%%%%%%%%%%%%%%%%%%%%%%%%%%%%%%%%%%%%%%%%%%%%%%%%%%%%%%%%%%%%%%%%%%%%%%%%
\begin{frontmatter}
%\title{Factorization technique and isochronous condition for coupled quadratic Li\'enard-type nonlinear systems}

%%%%%%%%%%%%%%%%%%%%%%%%%%%%%%%%%%%%%%%%%%%%%%%%%%%%%%%%%%%%%%%%%%%%%%%%%%%%%%%%%%%%%%%%%%%%%%%%%%%%%%%%%%%%%%%%%%%%%%%%%%%%%%%%%%%%%%%%%%%%%%%%%%%%%%%%%
\title{Factorization technique and isochronous condition for coupled quadratic and mixed Li\'enard-type nonlinear systems}
\author[rvt]{Ajey K. Tiwari}
\ead{ajey.nld@gmail.com}
\author[focal]{S. N. Pandey}
\ead{snp@mnnit.ac.in}
\author[vkc]{V. K. Chandrasekar\corref{cor1}}
\ead{chandru25nld@gmail.com}
\author[rvt]{M. Lakshmanan}
\ead{lakshman@cnld.bdu.ac.in}
\cortext[cor1]{Corresponding author}
\address[rvt]{Centre for Nonlinear Dynamics, School of Physics, Bharathidasan University, Tiruchirapalli - 620 024, India}
\address[focal]{Department of Physics, Motilal Nehru National Institute of Technology, Allahabad - 211 004, India}
\address[vkc]{Centre for Nonlinear Science and Engineering, School of Electrical and Electronics Engineering, SASTRA University, Thanjavur - 613 401, India}

%%%%%%%%%%%%%%%%%%%%%%%%%%%%%%%%%%%%%%%%%%%%%%%%%%%%%%%%%%%%%%%%%%%%%%%%%%%%%%%%%%%%%%%%%%%%%%%%%%%%%%%%%%%%%%%%%%%%%%%%%%%%%%%%%%%%%%%%%%%%%%%%%%%%%%%%%
\begin{abstract}
In this paper, we discuss a systematic and self consistent procedure to factorize a rather general class of coupled nonlinear ordinary differential equations (ODEs), namely coupled quadratic and mixed Li\'enard type equations, which include various physical and mathematical models. The procedure is broadly divided into two parts. In the first part, we consider a general factorized form for the equation under consideration in terms of some unknown functions and identify the determining equations for them. In the second part, we systematically solve the determining equations and identify the compatible factorizing form for this class of equations. In addition, we also discuss the problem of identification of isochronous dynamical systems belonging to the above class of equations. In particular, we deduce an isochronicity condition for the coupled quadratic Li\'enard equation. We also present specific examples of physical interest.
\end{abstract}

%%%%%%%%%%%%%%%%%%%%%%%%%%%%%%%%%%%%%%%%%%%%%%%%%%%%%%%%%%%%%%%%%%%%%%%%%%%%%%%%%%%%%%%%%%%%%%%%%%%%%%%%%%%%%%%%%%%%%%%%%%%%%%%%%%%%%%%%%%%%%%%%%%%%%%%%%

%\begin{keyword}
%nanoindentation \sep dislocations \sep plasticity \sep defects \sep crystal lattices \sep nucleation
%\PACS 82.40.Bj \sep 05.45.-a \sep 61.72.Bb %% MSC codes here, in the form: \MSC code \sep code
%\MSC 70K50 \sep 65P30 \sep 74A60 \sep 65Z05
%\end{keyword}

%%%%%%%%%%%%%%%%%%%%%%%%%%%%%%%%%%%%%%%%%%%%%%%%%%%%%%%%%%%%%%%%%%%%%%%%%%%%%%%%%%%%%%%%%%%%%%%%%%%%%%%%%%%%%%%%%%%%%%%%%%%%%%%%%%%%%%%%%%%%%%%%%%%%%%%%%

\end{frontmatter}

%%%%%%%%%%%%%%%%%%%%%%%%%%%%%%%%%%%%%%%%%%%%%%%%%%%%%%%%%%%%%%%%%%%%%%%%%%%%%%%%%%%%%%%%%%%%%%%%%%%%%%%%%%%%%%%%%%%%%%%%%%%%%%%%%%%%%%%%%%%%%%%%%%%%%%%%%
\section{Introduction}
The factorization method is well known in quantum mechanics for solving certain kind of ordinary differential equations (ODEs). It is an operational procedure which enables one to answer, in a direct manner, questions about a class of eigenvalue problems. The underlying idea is to consider a pair of first order differential-difference equations with boundary conditions \cite{inf,mie}. For example, decomposition of the quantum linear harmonic oscillator problem in terms of creation and annihilation operators is a case point. Recently, it has been shown by Rosu and his co-workers that at least in the case of some polynomial nonlinearities particular solutions may be found rather simply by an elegant method of factorizing them \cite{cor1,cor2}. This method has been explored in the case of scalar ODEs and nonlinear partial differential equations (PDEs) also and several classes of  solutions of many problems have been obtained rather straightforwardly \cite{cor3,leach,leach2,govinder}. Further, this method has been applied to the case of a system of coupled Li\'enard type equations with linear velocity terms and specific classes of Li\'enard type systems were identified for which particular solutions may be found by solving a Bernoulli equation \cite{haz}. 

Eventhough, this method plays a crucial role in understanding the nature of the various physical and mathematical models, it is very difficult to obtain the factorized form even in the case of scalar nonlinear ODEs. In the case of coupled ODEs the problem of factorizing the given equation becomes much more complex and one needs a systematic procedure to obtain the factorized form.  Our aim, in this paper, is to obtain the factorized form for a general class of coupled ODEs of the type (coupled Li\'enard type equations with quadratic velocities)
\begin{subequations}
\label{ch6.int4}
\begin{eqnarray}
\ddot{x}+h_1(x,y)\dot{x}^2+h_2(x,y)\dot{y}^2+h_3(x,y)\dot{x}\dot{y}+g_1(x,y)=0,\label{int4.1}  \\
\ddot{y}+h_4(x,y)\dot{x}^2+h_5(x,y)\dot{y}^2+h_6(x,y)\dot{x}\dot{y}+g_2(x,y)=0,\label{int4.2}
\end{eqnarray}
\end{subequations}
where ${h_i}^{'}s,\,i=1,....,6$ and ${g_j}^{'}s,\,j=1,2$ are functions of $x$ and $y$ and then extend this procedure to an even more general class of coupled mixed (quadratic and linear) Li\'enard type equations
\begin{subequations}
\label{int1}
\begin{eqnarray}
\hspace{-2cm}\ddot{x}+h_1(x,y)\dot{x}^2+h_2(x,y)\dot{y}^2+h_3(x,y)\dot{x}\dot{y}+f_1(x,y)\dot{x}+f_2(x,y)\dot{y}+g_1(x,y)=0,\label{int1.1}  \\
\hspace{-2cm}\ddot{y}+h_4(x,y)\dot{x}^2+h_5(x,y)\dot{y}^2+h_6(x,y)\dot{x}\dot{y}+f_3(x,y)\dot{x}+f_4(x,y)\dot{y}+g_2(x,y)=0.\label{int1.2}
\end{eqnarray}
\end{subequations}
The above equations include several physically and mathematically important equations and have been studied by many authors. In order to obtain the factorized forms for the above equations, in this paper we develop a systematic and self consistent procedure. For this purpose, we broadly divide our analysis into two parts. In the first part, we consider a general factorized form for Eq. (\ref{ch6.int4}) in the form of unknown functions to be determined. In fact an analysis of the scalar version of Eq. (\ref{ch6.int4}), namely
\begin{eqnarray}
\ddot{x}+h(x)\dot{x}^2+g(x)=0\label{int2}
\end{eqnarray}
can itself give important clues \cite{akt,nucci,muriel}. Using this knowledge, expanding and comparing the factorized equation with the original equation (\ref{ch6.int4}) we will get a set of PDEs for the coefficients ${h_i}^{'}s$ and ${g_j}^{'}s$ which in turn gives a set of determining equations for the unknown functions. To fix the factorized form corresponding to Eq. (\ref{ch6.int4}) we need to solve the obtained determining equations for the unknowns. Now, in the second part, we discuss the procedure to solve the set of determining equations consistently for the unknown functions. Solving the determining equations we can obtain the form of the unknowns which in turn will fix the factorized form corresponding to Eq. (\ref{ch6.int4}). To illustrate the effectiveness of this procedure we consider the coupled Mathews-Lakshmanan (ML) oscillator equations \cite{mathews,vkc4,cari,cari1,vkc3} and show how one can proceed systematically to identify the factorized form. It is to be noted that the study of Li\'enard type equation carried out by Hazra {\it et al.} \cite{haz} deals with identifying specific classes of Li\'enard type systems with linear velocity terms only for which particular solutions may be found by solving a Bernoulli equation. However, in this work, we focus our attention in developing a self consistent procedure in order to get the factorized form for the quadratic and mixed Li\'enard type equation. 

In addition to this, we will also discuss the isochronous properties associated with Eq. (\ref{ch6.int4}). For this purpose, we unearth an isochronicity condition for Eqs. (\ref{ch6.int4}) by transforming our system (Eq. (\ref{ch6.int4})) into a set of uncoupled simple harmonic oscillator equations as the latter set is a prototype of an isochronous system. The obtained isochronicity condition can be used to identify the class of equations exhibiting isochronous properties. We also consider a specific example exhibiting isochronous property. Finally, we include the linear velocity term in addition to quadratic term in Eq. (\ref{ch6.int4}) to get an overview of a more general class of equation, that is mixed Li\'enard type equation, Eq. (\ref{int1}). We show that the inclusion of the linear velocity term needs a small modification in the procedure discussed for the quadratic Li\'enard type equation.

The plan of the paper is as follows. In Sec. 2, we consider a general factorized form for Eq. (\ref{ch6.int4}) involving a set of unknown functions $\phi_k(x,y)$ and $\psi_{1,2}$, where $k=1,2,...,8$. We then develop a systematic algorithm for obtaining the factorized form in Sec. 3. Next, in Sec. 4, we systematically determine the forms of the unknown functions, ${\phi_k}^{'}s$, corresponding to the factorized form of Eq. (\ref{ch6.int4}). The forms of the functions $\psi_{1,2}$ are determined in Sec. 5. In Sec. 6, we demonstrate the procedure by considering the coupled ML oscillator equation. We discuss the isochronous property associated with Eq. (\ref{ch6.int4}), in Sec. 7. In Sec. 8, we consider a specific example corresponding to isochronous case. The case of mixed Li\'enard type equation is briefly discussed in Sec. 9. Factorization of Eq. (\ref{int2}) is discussed in Appendix A. In Appendix B, we discuss the factorization of scalar case corresponding to Eq. (\ref{int1}). Finally, our conclusions are given in Sec. 10.   

%%%%%%%%%%%%%%%%%%%%%%%%%%%%%%%%%%%%%%%%%%%%%%%%%%%%%%%%%%%%%%%%%%%%%%%%%%%%%%%%%%%%%%%%%%%%%%%%%%%%%%%%%%%%%%%%%%%%%%%%%%%%%%%%%%%%%%%%%%%%%%%%%%%%%%%%%%%%%%
\section{Factorization of the general case}
To start with, considering the scalar equation (\ref{int2}) it can be factorized as discussed in Appendix A. Taking this as a starting point, let us presume that the coupled quadratic Li\'enard type equation (\ref{ch6.int4}) can be factorized in the form
\begin{subequations}
\label{ch6.fact}
\begin{eqnarray}
&&[\phi_7(x,y)D-\phi_1(x,y)][\phi_5(x,y)D-\phi_2(x,y)]\psi_1(x,y)=0,\label{ch6.fact1}\\ 
&&[\phi_8(x,y)D-\phi_3(x,y)][\phi_6(x,y)D-\phi_4(x,y)]\psi_2(x,y)=0,\label{ch6.fact2}
\end{eqnarray}
\end{subequations}
where $D=\frac{d}{dt}$, ${\psi_{1,2}}^{'}s$ and ${\phi_k}^{'}s,\,k=1,2,...,8,$ are unknown functions of $x$ and $y$ to be determined. Now, the above set of equations can be rewritten as a set of first order coupled differential equations as 
\begin{subequations}
\label{ch6.solgen}
\begin{eqnarray}
&&[\phi_7(x,y)D-\phi_1(x,y)]P_1(x,y)=0,\\
&&[\phi_8(x,y)D-\phi_3(x,y)]P_2(x,y)=0,\\
&&[\phi_5(x,y)D-\phi_2(x,y)]\psi_1(x,y)=P_1(x,y),\\
&&[\phi_6(x,y)D-\phi_4(x,y)]\psi_2(x,y)=P_2(x,y).
\end{eqnarray}
\end{subequations}
Hence, our problem of finding the general solution of (\ref{ch6.int4}) is converted into simultaneously solving the above set of coupled first order differential equations, provided the decomposition (\ref{ch6.fact}) for (\ref{ch6.int4}) exists and can be explicitly found. A particular solution of Eq. (\ref{ch6.int4}) can be obtained by solving the reduced set of equations,
\begin{subequations}
\label{ch6.red}
\begin{eqnarray}
&&[\phi_5(x,y)D-\phi_2(x,y)]\psi_1(x,y)=0,\label{ch6.red1}\\
&&[\phi_6(x,y)D-\phi_4(x,y)]\psi_2(x,y)=0\label{ch6.red2},
\end{eqnarray}
\end{subequations}
which may in some cases be relatively simple.

Now, to identify the forms of the functions $h_i$ and $g_j,\,i=1,2,...,6$ and $j=1,2,$ for which Eq. (\ref{ch6.int4}) can be factorized in the form (\ref{ch6.fact}), we expand the latter and compare the resulting form with (\ref{ch6.int4}) appropriately for equivalence of (\ref{ch6.int4}) with (\ref{ch6.fact}). Then equating the governing powers of $\dot{x}$ and $\dot{y}$, we can show that the various coefficients $h_1,h_2,...,h_6$ and $g_1$ and $g_2$ are related to the unknown functions $\phi_1,\phi_2,...,\phi_8$ through the relations
\begin{subequations}
\label{ch6.hf}
\begin{eqnarray}
\hspace{-1.5cm}h_1&=\frac{1}{\delta}[\phi_6\phi_7\phi_8 \psi_{2y}(\psi_{1x}\phi_{5x}+\psi_{1xx}\phi_{5})-\phi_5\phi_7\phi_8 \psi_{1y}(\psi_{2x}\phi_{6x}+\psi_{2xx}\phi_{6})],\label{ch6.h1}\\
\hspace{-1.5cm}h_2&=\frac{1}{\delta}[\phi_6\phi_7\phi_8 \psi_{2y}(\psi_{1y}\phi_{5y}+\psi_{1yy}\phi_{5})-\phi_5\phi_7\phi_8 \psi_{1y}(\psi_{2y}\phi_{6y}+\psi_{2yy}\phi_{6})],\label{ch6.h2}
\end{eqnarray}
\begin{eqnarray}
\hspace{-1.5cm}h_3&=&\frac{1}{\delta}[\phi_6\phi_7\phi_8 \psi_{2y}(\psi_{1x}\phi_{5y}+2\phi_5\psi_{1xy}+\psi_{1y}\phi_{5x})\nonumber\\
\hspace{-1.4cm}&&-\phi_5\phi_7\phi_8 \psi_{1y}(\psi_{2x}\phi_{6y}+2\phi_6\psi_{2xy}+\psi_{2y}\phi_{6x})]\label{ch6.h3},\\
\hspace{-1.5cm}h_4&=&\frac{1}{\delta}[-\phi_6\phi_7\phi_8 \psi_{2x}(\psi_{1x}\phi_{5x}+\psi_{1xx}\phi_{5})+\phi_5\phi_7\phi_8 \psi_{1x}(\psi_{2x}\phi_{6x}+\psi_{2xx}\phi_{6})],\label{ch6.h4}\\
\hspace{-1.5cm}h_5&=&\frac{1}{\delta}[-\phi_6\phi_7\phi_8 \psi_{2x}(\psi_{1y}\phi_{5y}+\psi_{1yy}\phi_{5})+\phi_5\phi_7\phi_8 \psi_{1x}(\psi_{2y}\phi_{6y}+\psi_{2yy}\phi_{6})],\label{ch6.h5}\\
\hspace{-1.5cm}h_6&=&\frac{1}{\delta}[-\phi_6\phi_7\phi_8 \psi_{2x}(\psi_{1x}\phi_{5y}+2\phi_5\psi_{1xy}+\psi_{1y}\phi_{5x})\nonumber\\
\hspace{-1.4cm}&&+\phi_5\phi_7\phi_8 \psi_{1x}(\psi_{2x}\phi_{6y}+2\phi_6\psi_{2xy}+\psi_{2y}\phi_{6x})]\label{ch6.h6}
\end{eqnarray}
\end{subequations}
and the forms of ${g_j}^{'}s,\,j=1,2,$ as 
\begin{subequations}
\label{ch6.gf}
\begin{eqnarray}
g_1&=\frac{1}{\delta}\left[\phi_1\phi_2\phi_6\phi_8\psi_{2y}\psi_1 - \phi_3\phi_4\phi_5\phi_7\psi_{1y}\psi_2\right],  \label{ch6.gdef1}\\
g_2&=\frac{1}{\delta}\left[-\phi_1\phi_2\phi_6\phi_8\psi_{2x}\psi_1+\phi_3\phi_4\phi_5\phi_7\psi_{1x}\psi_2 \right], \label{ch6.gdef2}
\end{eqnarray}
\end{subequations}
where the quantity 
\begin{eqnarray}
\delta&=\phi_5\phi_6\phi_7\phi_8(\psi_{1x}\psi_{2y}-\psi_{2x}\psi_{1y})\neq 0.& \label{ch6.delta}
\end{eqnarray}
Further, as Eq. (\ref{ch6.int4}) does not contain the terms $\dot{x}$ and $\dot{y}$, their coefficients must be set equal to zero. Consequently we obtain the relations 
\begin{subequations}
\label{ch6.gft}
\begin{eqnarray}
 \phi_6\phi_8 \psi_{2y}[(\phi_7(\phi_2\psi_{1x}+\psi_{1}\phi_{2x})+\phi_1\phi_5\psi_{1x}]
=\phi_5\phi_7 \psi_{1y}[(\phi_8(\phi_4\psi_{2x}+\psi_{2}\phi_{4x})+\phi_3\phi_6\psi_{2x}],\label{ch6.f1}\\
 \phi_6\phi_8 \psi_{2y}[(\phi_7(\phi_2\psi_{1y}+\psi_{1}\phi_{2y})+\phi_1\phi_5\psi_{1y}]
=\phi_5\phi_7 \psi_{1y}[(\phi_8(\phi_4\psi_{2y}+\psi_{2}\phi_{4y})+\phi_3\phi_6\psi_{2y}],\label{ch6.f2}\\
 \phi_6\phi_8 \psi_{2x}[(\phi_7(\phi_2\psi_{1x}+\psi_{1}\phi_{2x})+\phi_1\phi_5\psi_{1x}]
=\phi_5\phi_7 \psi_{1x}[(\phi_8(\phi_4\psi_{2x}+\psi_{2}\phi_{4x})+\phi_3\phi_6\psi_{2x}],\label{ch6.f3}\\
 \phi_6\phi_8 \psi_{2x}[(\phi_7(\phi_2\psi_{1y}+\psi_{1}\phi_{2y})+\phi_1\phi_5\psi_{1y}]
=\phi_5\phi_7 \psi_{1y}[(\phi_8(\phi_4\psi_{2y}+\psi_{2}\phi_{4y})+\phi_3\phi_6\psi_{2y}].\label{ch6.f4}
\end{eqnarray}
\end{subequations}
Here, it is to be noted that one can also consider the equation including the terms $\dot{x}$ and $\dot{y}$, as in Eq. (\ref{int1}), that is the case of coupled mixed Li\'enard type equation. However, the above set of equations are the determining equations for the functions $\phi_2$ and $\phi_4$ (see Sec. 3 below). It is clear that considering the linear velocity term to be zero simplifies the above set of determining equations. If the linear velocity term is not zero then it corresponds to mixed Li\'enard type equation, Eq. (\ref{int1}), which will be discussed in Sec. 9.

Hence, we have factorized Eq. (\ref{ch6.int4}) in the form of Eq. (\ref{ch6.fact}). The connection between the coefficients of both the equations are given by the relations (\ref{ch6.hf})-(\ref{ch6.gft}). With the help of these relations one can obtain the factorized form by inverting the relations.

%%%%%%%%%%%%%%%%%%%%%%%%%%%%%%%%%%%%%%%%%%%%%%%%%%%%%%%%%%%%%%%%%%%%%%%%%%%%%%%%%%%%%%%%%%%%%%%%%%%%%%%%%%%%%%%%%%%%%%%%%%%%%%%%%%%%%%%%%%%%%%%%%%%%%%%%%%%%%%
\section{Systematic algorithm for obtaining the factorized form}
In order to get the form of the factorized equation (\ref{ch6.fact}) we have to solve the relations (\ref{ch6.hf})-(\ref{ch6.gft}). Here, in this section, we present a systematic algorithm to obtain the functions ${\phi_k}^{'}s$ in (\ref{ch6.fact}) by applying appropriate compatibility conditions on them.

Now, inverting the relations (\ref{ch6.gf}), one can get the forms of $\phi_1$ and $\phi_3$ as
\begin{eqnarray}
\phi_1=\frac{\phi_5\phi_7}{\phi_2}\bigg[\frac{g_1\psi_{1x}+g_2\psi_{1y}}{\psi_1} \bigg], \quad
\phi_3=\frac{\phi_6\phi_8}{\phi_4}\bigg[\frac{g_1\psi_{2x}+g_2\psi_{2y}}{\psi_2} \bigg]. \label{ch6.phi13gen2}
\end{eqnarray}
Multiplying Eq. (\ref{ch6.f1}) with $\psi_{1x}$ and Eq. (\ref{ch6.f3}) with $\psi_{1y}$ and adding, we get  
\begin{eqnarray}
\phi_6\phi_8(\phi_7(\psi_{1}\phi_{2x}+\psi_{1x}\phi_{2})+\phi_1 \phi_5\psi_{1x})(\psi_{1y}\psi_{2x}-\psi_{1x}\psi_{2y})=0.
\end{eqnarray}
Using $\phi_1$ from Eq. (\ref{ch6.phi13gen2}), we can rewrite the above equation in the form
\begin{eqnarray}
\phi_2\phi_{2x}\psi_1^2+\phi_2^2\psi_1\psi_{1x}+\phi_5^2\psi_{1x}(g_1\psi_{1x}+g_2\psi_{1y})=0.\label{ch6.p1}
\end{eqnarray}
Again multiplying Eq. (\ref{ch6.f1}) with $\psi_{2x}$ and Eq. (\ref{ch6.f3}) with $\psi_{2y}$ and adding, we get  
\begin{eqnarray}
\phi_4\phi_{4x}\psi_2^2+\phi_4^2\psi_2\psi_{2x}+\phi_6^2\psi_{2x}(g_1\psi_{2x}+g_2\psi_{2y})=0.\label{ch6.p3}
\end{eqnarray}
Similarly, doing the same for Eqs. (\ref{ch6.f2}) and (\ref{ch6.f4}), we arrive at the following relations,
\begin{subequations}
\label{ch6.p}
\begin{eqnarray}
&&\phi_2\phi_{2y}\psi_1^2+\phi_2^2\psi_1\psi_{1y}+\phi_5^2\psi_{1y}(g_1\psi_{1x}+g_2\psi_{1y})=0,\label{ch6.p2}\\
&&\phi_4\phi_{4y}\psi_2^2+\phi_4^2\psi_2\psi_{2y}+\phi_6^2\psi_{2y}(g_1\psi_{2x}+g_2\psi_{2y})=0.\label{ch6.p4}
\end{eqnarray}
\end{subequations}
Note that Eqs. (\ref{ch6.p1})-(\ref{ch6.p}) are effectively Riccati type equations for $\phi_2^2$ and $\phi_4^2$. With the help of Eqs. (\ref{ch6.h1}) and (\ref{ch6.h4}) one can get 
\begin{subequations}
\label{ch6.phi1}
\begin{eqnarray}
&&\phi_{5x}\psi_{1x}-(h_1\psi_{1x}+h_4\psi_{1y}-\psi_{1xx})\phi_{5}=0, \label{ch6.phi5.1}\\
&&\phi_{6x}\psi_{2x}-(h_1\psi_{2x}+h_4\psi_{2y}-\psi_{2xx})\phi_{6}=0,\label{ch6.phi6.1}
\end{eqnarray}
\end{subequations}
whereas from Eqs. (\ref{ch6.h2}) and (\ref{ch6.h5}) we get
\begin{subequations}
\label{ch6.phi2}
\begin{eqnarray}
&& \phi_{5y}\psi_{1y}-(h_2\psi_{1x}+h_5\psi_{1y}-\psi_{1yy})\phi_{5}=0, \label{ch6.phi5.2}\\
&& \phi_{6y}\psi_{2y}-(h_2\psi_{2x}+h_5\psi_{2y}-\psi_{2yy})\phi_{6}=0.\label{ch6.phi6.2}
\end{eqnarray}
\end{subequations}
Now, from Eqs. (\ref{ch6.h3}) and (\ref{ch6.h6}) one can get a set of PDEs for the functions $\phi_5$ and $\phi_6$ as
\begin{subequations}
\label{ch6.q}
\begin{eqnarray}
&&\psi_{1y}\phi_{5x}+\psi_{1x}\phi_{5y}+[2\psi_{1xy}-h_3\psi_{1x}-h_6\psi_{1y}]\phi_5=0, \label{ch6.q51}\\
&&\psi_{2y}\phi_{6x}+\psi_{2x}\phi_{6y}+[2\psi_{2xy}-h_3\psi_{2x}-h_6\psi_{2y}]\phi_6=0. \label{ch6.q52}
\end{eqnarray}
\end{subequations}
Substituting the values of $\phi_{5x},\,\phi_{5y},\,\phi_{6x}$ and $\phi_{6y}$ from Eqs. (\ref{ch6.phi1})-(\ref{ch6.phi2}) in Eqs. (\ref{ch6.q}), we get a set of PDEs for $\psi_1$ and $\psi_2$ as
\begin{subequations}
\label{ch6.psi}
\begin{eqnarray}
\hspace{-2cm}\psi_{1y}^2\psi_{1xx}+\psi_{1x}^2\psi_{1yy}-2\psi_{1x}\psi_{1y}\psi_{1xy}-h_2\psi_{1x}^3-h_4\psi_{1y}^3-(h_5-h_3)\psi_{1x}^2\psi_{1y}\nonumber\\
\hspace{1.2cm}-(h_1-h_6)\psi_{1x}\psi_{1y}^2=0,\label{ch6.psi1}\\
\hspace{-2cm}\psi_{2y}^2\psi_{2xx}+\psi_{2x}^2\psi_{2yy}-2\psi_{2x}\psi_{2y}\psi_{2xy}-h_2\psi_{2x}^3-h_4\psi_{2y}^3-(h_5-h_3)\psi_{2x}^2\psi_{2y}\nonumber\\
\hspace{1.2cm}-(h_1-h_6)\psi_{2x}\psi_{2y}^2=0.\label{ch6.psi2}
\end{eqnarray}
\end{subequations}
To get the factorized form for Eqs. (\ref{ch6.int4}), we need to solve the set of PDEs (\ref{ch6.psi}). Once we know the forms of $\psi_1$ and $\psi_2$ we can proceed further to obtain the ${\phi_k}^{'}s$. Now substituting the forms of $\psi_1$ and $\psi_2$ obtained by solving (\ref{ch6.psi}) into Eqs. (\ref{ch6.phi1})-(\ref{ch6.phi2}) one can easily get the values of $\phi_5$ and $\phi_6$ which on substitution in Eqs. (\ref{ch6.p}) gives the values of $\phi_2$ and $\phi_4$. Here, it is to be noted that the structure of the factorized form (\ref{ch6.fact}) suggests that one can always define $\tilde{\phi}_1=\frac{\phi_1}{\phi_7}$ and $\tilde{\phi}_3=\frac{\phi_3}{\phi_8}$. Now, we can rewrite Eq. (\ref{ch6.fact}) in terms of new functions $\tilde{\phi}_1$ and $\tilde{\phi}_3$ as 
\begin{subequations}
\label{ch7.fact}
\begin{eqnarray}
&&[D-\tilde{\phi}_1(x,y)][\phi_5(x,y)D-\phi_2(x,y)]\psi_1(x,y)=0,\label{ch7.fact1}\\ 
&&[D-\tilde{\phi}_3(x,y)][\phi_6(x,y)D-\phi_4(x,y)]\psi_2(x,y)=0.\label{ch7.fact2}
\end{eqnarray}
\end{subequations}
Then, the forms of $\tilde{\phi}_1$ and $\tilde{\phi}_3$ can be obtained from Eq. (\ref{ch6.phi13gen2}). Now, we can write the factorized form in terms of $\tilde{\phi}_1$ and $\tilde{\phi}_3$ (vide Eq. (\ref{ch7.fact})). Using the relations $\tilde{\phi}_1=\frac{\phi_1}{\phi_7}$ and $\tilde{\phi}_3=\frac{\phi_3}{\phi_8}$ the original factorized form, that is Eq. (\ref{ch6.fact}) can be obtained. In this way the factorization of Eq. (\ref{ch6.int4}) is complete. 

%%%%%%%%%%%%%%%%%%%%%%%%%%%%%%%%%%%%%%%%%%%%%%%%%%%%%%%%%%%%%%%%%%%%%%%%%%%%%%%%%%%%%%%%%%%%%%%%%%%%%%%%%%%%%%%%%%%%%%%%%%%%%%%%%%%%%%%%%%%%%%%%%%%%%%%%%%%%%%
\section{Determination of the functions ${\phi_k}^{'}s$}
In this section, we discuss the procedure to obtain the form of the functions $\psi_{1,2}$ and ${\phi_k}^{'}s,\,k=1,2,...,8,$ systematically. 

To start with, we integrate Eq. (\ref{ch6.phi5.1}) to obtain the form of $\phi_5$ as
\begin{eqnarray}
\phi_5=\frac{c_1(y)}{\psi_{1x}}\,e^{\int{\left(h_1+h_4\frac{\psi_{1y}}{\psi_{1x}}\right)dx}},\label{alt8}
\end{eqnarray}
where $c_1(y)$ is an arbitrary function of $y$. Similarly, solving Eq. (\ref{ch6.phi5.2}) we get yet another form of $\phi_5$ as
\begin{eqnarray}
\phi_5=\frac{c_2(x)}{\psi_{1y}}\,e^{\int{\left(h_5+h_2\frac{\psi_{1x}}{\psi_{1y}}\right)dy}},\label{alt9}
\end{eqnarray}
where $c_2(x)$ is an arbitrary functions of $x$. The relation between the functions $c_1$ and $c_2$ corresponding to Eqs. (\ref{alt8}) and (\ref{alt9}) can be determined by comparing both the forms of $\phi_5$. In doing so, we get
\begin{eqnarray}
\frac{c_1}{c_2}=\frac{\psi_{1x}}{\psi_{1y}}\,e^{\int{\left(h_5+h_2\frac{\psi_{1x}}{\psi_{1y}}\right)dy}-\int{\left(h_1+h_4\frac{\psi_{1y}}{\psi_{1x}}\right)dx}}.\label{alt10}
\end{eqnarray}
Hence, the form of $\phi_5$ can be obtained from Eqs. (\ref{alt8}) or (\ref{alt9}) provided $c_1$ and $c_2$ satisfy the relation (\ref{alt10}). Similarly, the form of $\phi_6$ can be obtained by solving Eqs. (\ref{ch6.phi6.1}) and (\ref{ch6.phi6.2}) as
\begin{eqnarray}
\phi_6&=\frac{c_3(y)}{\psi_{2x}}\,e^{\int{\left(h_1+h_4\frac{\psi_{2y}}{\psi_{2x}}\right)dx}},\label{alt11}
\end{eqnarray}
and 
\begin{eqnarray}
\phi_6&=\frac{c_4(x)}{\psi_{2y}}\,e^{\int{\left(h_5+h_2\frac{\psi_{2x}}{\psi_{2y}}\right)dy}},\label{alt12}
\end{eqnarray}
where $c_3(y)$ and $c_4(x)$ are arbitrary functions of $x$ and $y$, respectively. Here again, $c_3$ and $c_4$ are related by
\begin{eqnarray}
\frac{c_3}{c_4}=\frac{\psi_{2x}}{\psi_{2y}}\,e^{\int{\left(h_5+h_2\frac{\psi_{2x}}{\psi_{2y}}\right)dy}-\int{\left(h_1+h_4\frac{\psi_{2y}}{\psi_{2x}}\right)dx}}.\label{alt13}
\end{eqnarray}
Substituting the form of $\phi_5$ in Eqs. (\ref{ch6.p1}) and (\ref{ch6.p2}) and solving, we get the form of $\phi_2$ as
\begin{eqnarray}
\phi_2=\frac{\sqrt{2}}{\psi_1}\sqrt{c_5(y)-\int{\phi_5^2\psi_{1x}\left(g_1\psi_{1x}+g_2\psi_{1y}\right)dx}},\label{alt14}
\end{eqnarray}
and 
\begin{eqnarray}
\phi_2=\frac{\sqrt{2}}{\psi_1}\sqrt{c_6(x)-\int{\phi_5^2\psi_{1y}\left(g_1\psi_{1x}+g_2\psi_{1y}\right)dy}},\label{alt15}
\end{eqnarray}
where $c_5(y)$ and $c_6(x)$ are arbitrary function of $x$ and $y$, respectively, which are related by the relation
\begin{eqnarray}
\hspace{-1.8cm}c_5(y)-c_6(x)=\int{\phi_5^2\psi_{1x}\left(g_1\psi_{1x}+g_2\psi_{1y}\right)dx}-\int{\phi_5^2\psi_{1y}\left(g_1\psi_{1x}+g_2\psi_{1y}\right)dx}.\label{alt16}
\end{eqnarray}
Further, from Eqs. (\ref{ch6.p1}) and (\ref{ch6.p2}), one can check that the two expressions for $\phi_2$ should satisfy the compatibility criterion 
\begin{eqnarray}
\frac{\partial }{\partial y}\left[\phi_5^2\psi_{1x}(g_1\psi_{1x}+g_2\psi_{1y})\right]=\frac{\partial }{\partial x}\left[\phi_5^2\psi_{1y}(g_1\psi_{1x}+g_2\psi_{1y})\right].\label{alt7.1}
\end{eqnarray}

Similarly, the form of $\phi_4$ can be written with the help of Eqs. (\ref{ch6.p3}) and (\ref{ch6.p4}) as
\begin{eqnarray}
\phi_4=\frac{\sqrt{2}}{\psi_2}\sqrt{c_7(y)-\int{\phi_6^2\psi_{2x}\left(g_1\psi_{2x}+g_2\psi_{2y}\right)dx}},\label{alt18}
\end{eqnarray}
and 
\begin{eqnarray}
\phi_4=\frac{\sqrt{2}}{\psi_2}\sqrt{c_8(x)-\int{\phi_6^2\psi_{2y}\left(g_1\psi_{2x}+g_2\psi_{2y}\right)dy}},\label{alt19}
\end{eqnarray}
where the arbitrary functions $c_7(y)$ and $c_8(x)$ are related by the expression 
\begin{eqnarray}
\hspace{-1.8cm}c_7(y)-c_8(x)=\int{\phi_6^2\psi_{2x}\left(g_1\psi_{2x}+g_2\psi_{2y}\right)dx}-\int{\phi_6^2\psi_{2y}\left(g_1\psi_{2x}+g_2\psi_{2y}\right)dx}\label{alt20}
\end{eqnarray}
and $\phi_4$ should satisfy the compatibility criterion (as may be seen from Eqs. (\ref{ch6.p3}) and (\ref{ch6.p4}))
\begin{eqnarray}
\frac{\partial }{\partial y}\left[\phi_6^2\psi_{2x}(g_1\psi_{2x}+g_2\psi_{2y})\right]=\frac{\partial }{\partial x}\left[\phi_6^2\psi_{2y}(g_1\psi_{2x}+g_2\psi_{2y})\right].\label{alt7.2}
\end{eqnarray}

Now, the forms of the functions $\tilde{\phi}_1$ and $\tilde{\phi}_3$  can be obtained from Eqs. (\ref{ch6.phi13gen2}) as
\begin{eqnarray}
\tilde{\phi}_1=\frac{c_1(g_1\psi_{1x}+g_2\psi_{1y})}{\sqrt{2}\psi_{1x}}\frac{e^{\int{\left(h_1+h_4\frac{\psi_{1x}}{\psi_{1y}}\right)dx}}}{\sqrt{c_5(y)-\int{\phi_5^2\psi_{1x}\left(g_1\psi_{1x}+g_2\psi_{1y}\right)dx}}},\label{alt22}
\end{eqnarray}
and
\begin{eqnarray}
\tilde{\phi}_3=\frac{c_4(g_1\psi_{2x}+g_2\psi_{2y})}{\sqrt{2}\psi_{2x}}\frac{e^{\int{\left(h_1+h_4\frac{\psi_{2x}}{\psi_{2y}}\right)dx}}}{\sqrt{c_7(y)-\int{\phi_6^2\psi_{2x}\left(g_1\psi_{2x}+g_2\psi_{2y}\right)dx}}}.\label{alt23}
\end{eqnarray}
Now, we know the forms of ${\phi_k}^{'}s$ in terms of the ${h_i}^{'}s,\,{g_j}^{'}s$ and $\psi_{1,2}$ where the only unknown terms are $\psi_{1,2}$. So the problem of factorizing the system under consideration reduces to determining the suitable forms of $\psi_{1,2}$ which will be discussed in the next section.

%%%%%%%%%%%%%%%%%%%%%%%%%%%%%%%%%%%%%%%%%%%%%%%%%%%%%%%%%%%%%%%%%%%%%%%%%%%%%%%%%%%%%%%%%%%%%%%%%%%%%%%%%%%%%%%%%%%%%%%%%%%%%%%%%%%%%%%%%%%%%%%%%%%%%%%%%%%%%%
\section{Determination of the functions $\psi_{1,2}$}
In this section, we discuss the procedure to identify suitable forms for the functions $\psi_{1,2}$, so that the factorization can be completed. For this purpose, we proceed as follows. 

It is always possible to rewrite the given set of coefficients ${h_i}^{'}s$ in (\ref{ch6.int4}) with a common denominator. Hence, we define
\begin{eqnarray}
h_i=\frac{\kappa_i(x,y)}{G(x,y)^p},\quad i=1,2,...,8,\label{alt1}
\end{eqnarray}
where $\kappa_i(x,y)$ and $G(x,y)$ are functions of $x$ and $y$ and $p$ is an arbitrary parameter. Now, to obtain the forms of $\psi_1$ and $\psi_2$ we need to solve the set of coupled PDEs (\ref{ch6.psi}). For this purpose, we need an ansatz for the functions $\psi_1$ and $\psi_2$ as it is very difficult to solve otherwise. One can consider the form of $\psi_{1,2}$ as a rational one where the numerator and denominator are both functions of $x$ and $y$. Now, as the coefficients ${h_i}^{'}s$ are also of rational form, we consider the denominator of $\psi_{1,2}$ as $G(x,y)$. Since $\psi_{1,2}$ is in rational form while taking differentiation or integration the form of the denominator remains the same but the power of the denominator decreases or increases by a unit order from that of the initial one. Thus, instead of considering the denominator to be just of the form $G(x,y)$ one can consider a more general form as $G^q(x,y)$, where $q$ is an arbitrary real number. Hence the anasatz for $\psi_{1,2}$ can be considered as
\begin{eqnarray}
\psi_{1,2}=\frac{F_{1,2}(x,y)}{G^q(x,y)},\label{alt2}
\end{eqnarray}
where $F_{1,2}(x,y)$ are functions to be determined. Substituting the above forms of $\psi_{1}$ and $h_i^{'}s$ in Eq. (\ref{ch6.psi1}) and simplifying the latter, we get
\begin{eqnarray}
G^{p}\left\{[D_y(F_1G^{q})]^2[D_x^2(F_1G^{q})]+[D_x(F_1G^{q})]^2[D_y^2(F_1G^{q})]-2[D_x(F_1G^{q})][D_y(F_1G^{q})]\right.\nonumber\\
\hspace{-.8cm}\left.\times [D_xD_y(F_1G^{q})]\right\}-2qF_1G^{p+q-2}\left\{[D_x(G)]^2+[D_y(G)]^2-2[D_x(G)][D_y(G)]\right.\nonumber\\
\left.-G\left(D_x^2(G)+D_y^2(G)-2D_xD_y(G)\right)\right\}-\kappa_2[D_x(F_1G^{q})]^3-\kappa_4[D_y(F_1G^{q})]^3\,\,\,\,\,\,\,\,\,\,\nonumber\\
+(\kappa_1-\kappa_6)[D_x(F_1G^{q})][D_y(F_1G^{q})]^2-(\kappa_5-\kappa_3)[D_y(F_1G^{q})][D_x(F_1G^{q})]^2=0,\label{alt6}
\end{eqnarray}
and identically for $F_2$, 
%\begin{eqnarray}G\{[D_y(F_2G^{-q})]^2D_x^2(F_2G^{-q})+[D_x(F_2G^{-q})]^2D_y^2(F_2G^{-q})+2D_x(F_2G^{-q})D_y(F_2G^{-q})\nonumber\\
%\hspace{-.8cm}\times D_xD_y(F_2G^{-q})\}-\kappa_4[D_y(F_2G^{-q})]^3-(\kappa_5-\kappa_3)D_y(F_2G^{-q})[D_x(F_2G^{-q})]^2\nonumber\\
%-2qG^{3q-2}\{(D_x(G))^2+(D_y(G))^2+D_x(G)D_y(G)\}-\kappa_2[D_x(F_2G^{-q})]^3\nonumber\\
%\qquad \qquad \qquad+(\kappa_1-\kappa_6)D_x(F_2G^{-q})[D_y(F_2G^{-q})]^2=0,\label{alt7}
%\end{eqnarray}
where the Hirota's $D-$operator is defined as
\begin{equation}
D_x^n\,(f.g)=(\partial_{x_1}-\partial_{x_2})^n f(x_1)g(x_2)\mid_{x_2=x_1=x}.\label{alt5}
\end{equation}
As we know the form of the function $G$, we need only to determine the forms of the functions $F_{1,2}$. For obtaining admissible forms of $\psi_{1,2}$ we need to choose suitable anasatz for $F_{1,2}$. For example, we can consider $F_{1,2}$ as polynomials in $x$ and $y$. Substituting the polynomial forms of $F_{1,2}$ into their determining equations (\ref{alt6}) and the identical equation for $F_2$, and equating the various functions of $x$ and $y$ to zero we get a set of resultant determining equations. Solving the obtained resultant equations we can determine the suitable forms of $F_{1,2}$ which in turn fixes the form of $\psi_{1,2}$. Hence, we can conclude that in principle it is always possible to get the factorized form for Eq. (\ref{ch6.int4}) for all the forms of the functions $h_i$ and $g_j$, provided ${F_{1,2}}^{'}s$ can be found. 

%%%%%%%%%%%%%%%%%%%%%%%%%%%%%%%%%%%%%%%%%%%%%%%%%%%%%%%%%%%%%%%%%%%%%%%%%%%%%%%%%%%%%%%%%%%%%%%%%%%%%%%%%%%%%%%%%%%%%%%%%%%%%%%%%%%%%%%%%%%%%%%%%%%%%%%%%%%%%%
\section{Example:}
In this section, we demonstrate the effectiveness of the procedure discussed in the previous sections by considering an example of this class and then factorizing it properly. We consider the general form of coupled Mathews-Lakshmanan oscillators \cite{cari,cari1,vkc3,vkc2}, that is 
\begin{subequations}
\label{ch6.1ml}
\begin{eqnarray}
\ddot{x}-\frac{\lambda x(1+\lambda y^2)\dot{x}^2+\lambda x(1+\lambda x^2)\dot{y}^2-2\lambda^2x^2y\dot{x}\dot{y}-\alpha_1x}{1+\lambda r^2}=0,\label{ch6.ml1}\\
\ddot{y}-\frac{\lambda y(1+\lambda y^2)\dot{x}^2+\lambda y(1+\lambda x^2)\dot{y}^2-2\lambda^2xy^2\dot{x}\dot{y}-\alpha_2y}{1+\lambda r^2}=0,\label{ch6.ml2}
\end{eqnarray}
\end{subequations}
where $r^2=x^2+y^2$ and $\lambda, \alpha_1$ and $\alpha_2$ are arbitrary constants. 

%%%%%%%%%%%%%%%%%%%%%%%%%%%%%%%%%%%%%%%%%%%%%%%%%%%%%%%%%%%%%%%%%%%%%%%%%%%%%%%%%%%%%%%%%%%%%%%%%%%%%%%%%%%%%%%%%%%%%%%%%%%%%%%%%%%%%%%%%%%%%%%%%%%%%%%%%%%%%%
\subsection{Factorization of Coupled ML oscillator equation}
As a first step, we compare Eq. (\ref{ch6.1ml}) with Eq. (\ref{ch6.int4}) to obtain the forms of the functions ${h_i}^{'}s$ and ${g_j}^{'}s$ as
\begin{eqnarray}
h_1=-\frac{\lambda x(1+\lambda y^2)}{1+\lambda r^2},\,h_2=-\frac{\lambda x(1+\lambda x^2)}{1+\lambda r^2},\,h_3=\frac{2\lambda^2 x^2y}{1+\lambda r^2},\,g_1=\frac{\alpha_1x}{1+\lambda r^2},\nonumber\\
h_4=-\frac{\lambda y(1+\lambda y^2)}{1+\lambda r^2},\,h_5=-\frac{\lambda y(1+\lambda x^2)}{1+\lambda r^2},\,h_6=\frac{2\lambda^2 xy^2}{1+\lambda r^2},\,g_2=\frac{\alpha_2y}{1+\lambda r^2}.\label{ch6.ml3}
\end{eqnarray}

It is clear from Eq. (\ref{ch6.ml3}) that all the ${h_i}^{'}s$ have a common denominator. Hence, we can construct suitable forms of $\psi_{1,2}$ with the help of the procedure discussed in the previous sections. For this, we consider the function $F_{1,2}(x,y)$ as polynomial functions in $x$ and $y$ as
\begin{equation}
F_{1,2}(x,y)=a_{1,2}x+b_{1,2}y,\label{ch6.ml4}
\end{equation}
where $a_{1,2}$ and $b_{1,2}$ are arbitrary parameters. Now, substituting this form of $F_{1,2}$ in Eq. (\ref{alt6}) for $F_1$ and its counterpart for $F_2$ and equating the various coefficients of the independent parameters to zero we get a set of algebraic equations in $a_{1,2},\,b_{1,2},\,p$ and $q$. Solving these equations consistently, we get
\begin{equation}
 a_{1,2}={arbitrary},\,b_{1,2}={arbitrary},\,p=1,\, q=\frac{1}{2}.\label{ch6.ml5}
\end{equation}
Thus, we find that the following forms of $\psi_{1,2}$ are compatible,
\begin{eqnarray}
\psi_1=\frac{x}{\sqrt{1+\lambda r^2}},\,\,\psi_2=\frac{y}{\sqrt{1+\lambda r^2}}.\label{ch6.ml6}
\end{eqnarray}
It is to be noted that the above forms of $\psi_{1,2}$ also satisfy the compatibility criteria given by Eqs. (\ref{alt7.1}) and (\ref{alt7.2}). Hence we can now determine the forms of $\phi_5$ and $\phi_6$. For this purpose, we substitute Eq. (\ref{ch6.ml6}) into Eqs. (\ref{alt8}) and (\ref{alt9}) and solving them consistently, we get the form of $\phi_5$ as
\begin{equation}
\phi_5=1+\lambda r^2,  \label{ch6.ml7}
\end{equation}
where the relation between $c_1(y)$ and $c_2(x)$ from Eq. (\ref{alt10}) turns out to be
\begin{eqnarray}
\frac{c_1}{c_2}=-\frac{1+\lambda y^2}{\lambda x}.\label{c1}
\end{eqnarray}
As $c_1$ is  function of $y$ only and $c_2$ is a function of $x$ only, we can identify $c_1(y)=1+\lambda y^2$ and $c_2(x)=-\lambda x$. Similarly, the form of $\phi_6$ can be obtained from Eqs. (\ref{alt11}) and (\ref{alt12}) as
\begin{equation}
\phi_6=1+\lambda r^2,  \label{ch6.ml7.1}
\end{equation}
where from Eq. (\ref{alt13}) we have $\frac{c_3}{c_4}=-\frac{\lambda y}{1+\lambda x^2}$. Hence, we can fix $c_3(y)=-\lambda y$ and $c_4(x)=1+\lambda x^2$ as $c_3$ is a function of $y$ alone, whereas $c_4$ is a function of $x$ alone. 

Now, to get the form of $\phi_2$ and $\phi_4$ we first check the compatibility conditions given by Eqs. (\ref{alt7.1}) and (\ref{alt7.2}). Doing so we have to necessarily fix $\alpha _1=\alpha _2=\alpha$ in Eq. (\ref{ch6.ml3}). Now, the function $\phi_2$ can be obtained from Eqs. (\ref{alt14}) or (\ref{alt15}) as
\begin{equation}
\phi_2=\pm\sqrt{-\alpha},\label{ch6.ml10}
\end{equation}
where the relation between the functions $c_5$ and $c_6$ turns out to be $c_5-c_6=-\frac{\alpha}{2\lambda}$. The relation between $c_5$ and $c_6$ suggest various possibilities for their forms. However, we need to find $c_5$ and $c_6$ such that they are consistent with Eqs. (\ref{alt14}) and (\ref{alt15}). For example, one can consider three simplest forms as $(i)\,c_5=-\frac{\alpha}{2\lambda}$ and $c_6=0$, $(ii)\,c_5=0$ and $c_6=-\frac{\alpha}{2\lambda}$ and $(iii)\,c_5=-\frac{\alpha}{4\lambda}$ and $c_6=\frac{\alpha}{4\lambda}$. One can check that only case $(i)$ is consistent with Eqs. (\ref{alt14}) and (\ref{alt15}). Similarly, we can get the form of $\phi_4$ as
\begin{equation}
\phi_4=\pm\sqrt{-\alpha},\label{ch6.ml10.1}
\end{equation}
where $c_7(y)=0$ and $c_8(x)=-\frac{\alpha}{2\lambda}$. Here also we need to fix the forms of $c_7$ and $c_8$ so that they are consistent with Eqs. (\ref{alt18}) and (\ref{alt19}). One can check that the other obvious choices, that is $c_7(y)=-\frac{\alpha}{2\lambda}$ and $c_8(x)=0$ and $c_5=-\frac{\alpha}{4\lambda}$ and $c_6=\frac{\alpha}{4\lambda}$ are not consistent with Eqs. (\ref{alt18}) and (\ref{alt19}).

Finally, the forms of $\tilde{\phi}_1$ and $\tilde{\phi}_3$ can be obtained with the help of Eqs. (\ref{alt22}) and (\ref{alt23}) as
\begin{equation}
\tilde{\phi}_1=\tilde{\phi}_3=\mp\frac{\sqrt{-\alpha}}{1+\lambda r^2}.\label{ch6.3ml7}
\end{equation}
Now, we know the forms of all the functions ${\phi_k}^{'}s,\,k=1,2,...,8,$ which completes the factorization of the coupled ML oscillator. The factorized form can be then written as
\begin{subequations}
\label{ch6.2ml}
\begin{eqnarray}
&&\left[(1+\lambda r^2)D\pm\sqrt{-\alpha}\right]\left[(1+\lambda r^2)D\mp\sqrt{-\alpha}\right]\,\frac{x}{\sqrt{1+\lambda r^2}}=0,\label{ch6.2ml1}\\
&&\left[(1+\lambda r^2)D\pm\sqrt{-\alpha}\right]\left[(1+\lambda r^2)D\mp\sqrt{-\alpha}\right]\,\frac{y}{\sqrt{1+\lambda r^2}}=0.\label{ch6.2ml2}
\end{eqnarray}
\end{subequations}

%%%%%%%%%%%%%%%%%%%%%%%%%%%%%%%%%%%%%%%%%%%%%%%%%%%%%%%%%%%%%%%%%%%%%%%%%%%%%%%%%%%%%%%%%%%%%%%%%%%%%%%%%%%%%%%%%%%%%%%%%%%%%%%%%%%%%%%%%%%%%%%%%%%%%%%%%%%%%%
\subsection{Integrability of coupled ML oscillator equation}
The above factorized form can be used to obtain the general solution of the coupled ML oscillator. For this purpose we can write Eq. (\ref{ch6.2ml1}) with the choice of both the signs as
\begin{subequations}
\label{ch6.1sol}
\begin{eqnarray}
&&\left[(1+\lambda r^2)D+\sqrt{-\alpha}\right]\left[(1+\lambda r^2)D-\sqrt{-\alpha}\right]\,\frac{x}{\sqrt{1+\lambda r^2}}=0,\label{ch6.1sol1}\\
&&\left[(1+\lambda r^2)D-\sqrt{-\alpha}\right]\left[(1+\lambda r^2)D+\sqrt{-\alpha}\right]\,\frac{x}{\sqrt{1+\lambda r^2}}=0.\label{ch6.1sol2}
\end{eqnarray}
\end{subequations}
Considering $\tilde{D}=(1+\lambda r^2)D$, the above Eqs. (\ref{ch6.1sol}) can be written in the form of the harmonic oscillator like equation,
\begin{subequations}
\label{ch6.111sol}
\begin{eqnarray}
&&\left[\tilde{D}+\sqrt{-\alpha}\right]\left[\tilde{D}-\sqrt{-\alpha}\right]\,\frac{x}{\sqrt{1+\lambda r^2}}=0,\label{ch6.111sol1}\\
&&\left[\tilde{D}-\sqrt{-\alpha}\right]\left[\tilde{D}+\sqrt{-\alpha}\right]\,\frac{x}{\sqrt{1+\lambda r^2}}=0.\label{ch6.111sol2}
\end{eqnarray}
\end{subequations}
One can always rewrite Eqs. (\ref{ch6.111sol}) in the form
\begin{eqnarray}
\frac{\tilde{D}\left[\tilde{D}[\frac{x}{\sqrt{1+\lambda r^2}}]-\sqrt{-\alpha}\frac{x}{\sqrt{1+\lambda r^2}}\right]}
{\tilde{D}[\frac{x}{\sqrt{1+\lambda r^2}}]-\sqrt{-\alpha}\frac{x}{\sqrt{1+\lambda r^2}}}
+\frac{\tilde{D}\left[\tilde{D}[\frac{x}{\sqrt{1+\lambda r^2}}]+\sqrt{-\alpha}\frac{x}{\sqrt{1+\lambda r^2}}\right]}
{\tilde{D}[\frac{x}{\sqrt{1+\lambda r^2}}]+\sqrt{-\alpha}\frac{x}{\sqrt{1+\lambda r^2}}}=0.\label{ch6.2sol}
\end{eqnarray}
Integrating Eq. (\ref{ch6.2sol}) once, we arrive at
\begin{eqnarray}
I_1=\left(\tilde{D}[\frac{x}{\sqrt{1+\lambda r^2}}]+\sqrt{-\alpha}\frac{x}{\sqrt{1+\lambda r^2}}\right) 
\left(\tilde{D}[\frac{x}{\sqrt{1+\lambda r^2}}]-\sqrt{-\alpha}\frac{x}{\sqrt{1+\lambda r^2}}\right),\label{ch6.3sol}
\end{eqnarray}
where $I_1$ is an integration constant. Now, (\ref{ch6.3sol}) can be rewritten as
\begin{eqnarray}
I_1=\left(\tilde{D}[\frac{x}{\sqrt{1+\lambda r^2}}]\right)^2+\alpha \left(\frac{x}{\sqrt{1+\lambda r^2}}\right)^2.\label{ch6.4sol}
\end{eqnarray}
Similarly, one can write from Eq. (\ref{ch6.2ml2}) as
\begin{eqnarray}
I_2=\left(\tilde{D}[\frac{y}{\sqrt{1+\lambda r^2}}]\right)^2+\alpha \left(\frac{y}{\sqrt{1+\lambda r^2}}\right)^2,\label{ch6.5sol}
\end{eqnarray}
where $I_2$ is a constant of integration. Now, substituting $\tilde{D}=(1+\lambda r^2)D$ and simplifying Eqs. (\ref{ch6.4sol}) and (\ref{ch6.5sol}), we get
\begin{eqnarray}
\left(\dot{x}(1+\lambda y^2)-\lambda xy \dot{y}\right)^2+\alpha x^2=I_1(1+\lambda r^2),\label{ch6.6sol}
\end{eqnarray}
and \begin{eqnarray}
\left(\dot{y}(1+\lambda x^2)-\lambda xy \dot{x}\right)^2+\alpha y^2=I_2(1+\lambda r^2).\label{ch6.7sol}
\end{eqnarray}
To deduce the third integration constant we use the first integral of harmonic oscillator equation
\begin{equation}
 I_3=\psi_2\tilde{D}\psi_1- \psi_1\tilde{D}\psi_2,\label{ch6.9sol}
\end{equation}
where $\tilde{D}=(1+\lambda r^2)\frac{d}{dt}$. Using the forms of $\psi_1,\psi_2$ and $\tilde{D}$ in Eq. (\ref{ch6.9sol}) and simplifying, we get the 
form of the integration constant $\tilde{I}_2$ as
\begin{equation}
 I_3=\dot{x}y-x\dot{y}.\label{ch6.11sol}
\end{equation}

Adding Eqs. (\ref{ch6.6sol}) and (\ref{ch6.7sol}) and simplifying, we arrive at
\begin{eqnarray}
\tilde{I}_1=\frac{\alpha r^2+\dot{x}^2\left(\left(1+\lambda y^2\right)^2+\lambda^2x^2y^2\right)+\dot{y}^2\left(\left(1+\lambda x^2\right)^2+\lambda^2x^2y^2\right)-2\lambda xy\dot{x}\dot{y}\left(2+\lambda r^2\right)}{1+\lambda r^2},\nonumber\\
\label{ch6.8sol}
\end{eqnarray}
where $\tilde{I}_1=I_1+I_2$ is a new integration constant. Once we know the form of the two integrals of motion $\tilde{I}_1$ and $I_3$, we can deduce the linearizing transformations with the help of the procedure discussed by Chandrasekar {\it et al.} \cite{vkc3}. For this purpose, we consider the first integrals $\tilde{I_1}$ and $I_3$. Now, rewriting the first integrals in the form
\begin{eqnarray}
\tilde{I}_1&=&\frac{\alpha r^2+\dot{x}^2\left(\left(1+\lambda y^2\right)^2+\lambda^2x^2y^2\right)+\dot{y}^2\left(\left(1+\lambda x^2\right)^2+\lambda^2x^2y^2\right)-2\lambda xy\dot{x}\dot{y}\left(2+\lambda r^2\right)}{2\lambda (x\dot{x}+y\dot{y})}\nonumber\\
&&\times\frac{d}{dt}\log{(1+\lambda r^2)}=\frac{dw_1}{dz_1},\label{lin011}\\
I_3&=&{y^2}\frac{d}{dt}\left(\frac{x}{y}\right)=\frac{dw_2}{dz_2},\label{lin01}
\end{eqnarray}
we identify the following set of linearizing transformations
\begin{eqnarray}
w_1&=&\log(1+\lambda r^2),\qquad \qquad  w_2=\frac{x}{y},\nonumber\\
z_1&=&\int{\frac{2\lambda (x\dot{x}+y\dot{y})\,dt}{\alpha r^2+\dot{x}^2\left(\left(1+\lambda y^2\right)^2+\lambda^2x^2y^2\right)+\dot{y}^2\left(\left(1+\lambda x^2\right)^2+\lambda^2x^2y^2\right)-2\lambda xy\dot{x}\dot{y}\left(2+\lambda r^2\right)}},\nonumber\\
z_2&=&\int{\frac{dt}{y^2}}.\label{ch6.12sol}
\end{eqnarray}
Rewriting the first integrals $\tilde{I}_1$ and $I_3$ in the integral form and identifying them in terms of the new variables, we get 
$w_1=\tilde{I}_1z_1$ and $w_2=I_3z_2$. From this one can get the relation between the variables $x$ and $y$ with $z_1$ and $z_2$, respectively, 
(the integration constant is fixed to be zero without loss of generality) as
\begin{equation}
1+\lambda r^2=e^{\tilde{I}_1z_1} \quad {and} \quad x=I_3z_2y.\label{ch6.13sol}
\end{equation}
Making use of Eqs. (\ref{ch6.8sol}), (\ref{ch6.11sol}) and (\ref{ch6.12sol}), we can write 
\begin{equation}
 dz_1=\frac{2\lambda \sqrt{\left(\tilde{I}_1-I_3^2\left(1+\lambda r^2\right)\right)\left(1+\lambda r^2\right)-\alpha r^4}}{I_1(1+\lambda r^2)}dt.\label{ch6.14sol}
\end{equation}
Using the result $1+\lambda r^2=e^{\tilde{I}_1z_1}$ the above expression can be rewritten as
\begin{eqnarray}
 dz_1=\frac{2\lambda}{\tilde{I}_1}\sqrt{\left(\tilde{I}_1+\frac{2\alpha }{\lambda^2}\right) e^{-\tilde{I}_1z_1}-\left(I_3^2+\frac{\alpha }{\lambda^2}\right)-\frac{\alpha }{\lambda^2} e^{-2\tilde{I}_1z_1}}dt.\label{ch6.15sol}
\end{eqnarray}
To get the third integration constant we integrate the above equation. Doing this, we get
\begin{eqnarray}
I_4-t=\frac{1}{2\lambda \sqrt{I_3^2+\frac{\alpha }{\lambda^2}}} \tan^{-1}\left[\frac{2\left(I_3^2+\frac{\alpha }{\lambda^2}\right)-\left(\tilde{I}_1+\frac{2\alpha }{\lambda^2} \right) e^{-\tilde{I}_1z_1}}{2\sqrt{I_3^2+\frac{\alpha }{\lambda^2}}\sqrt{\left(\tilde{I}_1+\frac{2\alpha }{\lambda^2} \right) e^{-\tilde{I}_1z_1}-\left(I_3^2+\frac{\alpha }{\lambda^2}\right)-\frac{\alpha}{\lambda^2} e^{-2\tilde{I}_1z_1}
}}\right], \label{ch6.16sol}
\end{eqnarray}
where $I_4$ is the fourth integral of motion. Now making use these four integrals of motion, namely (\ref{ch6.6sol}), (\ref{ch6.7sol}), (\ref{ch6.11sol}) and (\ref{ch6.16sol}), the general solution can be straightforwardly constructed. The resultant solution also agrees with Eq. (5.40) of Chandrasekar {\it et al.} \cite{vkc3}.

%%%%%%%%%%%%%%%%%%%%%%%%%%%%%%%%%%%%%%%%%%%%%%%%%%%%%%%%%%%%%%%%%%%%%%%%%%%%%%%%%%%%%%%%%%%%%%%%%%%%%%%%%%%%%%%%%%%%%%%%%%%%%%%%%%%%%%%%%%%%%%%%%%%%%%%%%%%%%%
\section{Isochronous condition}
In the previous sections, we paid our attention to get the factorized form of Eq. (\ref{ch6.int4}) systematically. Now, in this part of the paper 
we are interested in identifying the form of the equation belonging to Eq. (\ref{ch6.int4}) which exhibits isochronous properties. For this purpose, we 
transform our system (Eq. (\ref{ch6.int4})) into a set of uncoupled simple harmonic oscillator equations as the latter ones are prototypes of isochronous systems. 

Let us consider a system uncoupled harmonic oscillator equations of the form
\begin{subequations}
\label{shm}
\begin{eqnarray}
\ddot{\psi_1}+\omega_1\psi_1=0,\label{shm1}\\
\ddot{\psi_2}+\omega_2\psi_2=0.\label{shm2}
\end{eqnarray}
\end{subequations}
Eqs. (\ref{shm}) can be rewritten in the form of factorized equations as
\begin{subequations}\label{ch6.free}
\begin{eqnarray}
&&[D\pm\sqrt{-\omega_{1}}][D\mp\sqrt{-\omega_{1}}]\psi_1=0,\\
&&[D\pm\sqrt{-\omega_{2}}][D\mp\sqrt{-\omega_{2}}]\psi_2=0,\label{ch6.iso1}
\end{eqnarray}
\end{subequations}
where $\omega_1 $ and $\omega_2$ are constants and $\psi_1$ and $\psi_2$ are eigen functions. If system (\ref{ch6.fact}) exhibits isochronous property then it can be obtained from Eq. (\ref{ch6.free}) with the help of a suitable transformation. Hence, choosing the form of the functions $\psi_i$ appropriately one can transform Eq. (\ref{ch6.free}) to Eq. (\ref{ch6.fact}). Then one can easily identify the forms of the functions in (\ref{ch6.fact}) as $\phi_1=\phi_2=\sqrt{-\omega_1},\,\phi_3=\phi_4=\sqrt{-\omega_2}$ and $\phi_5=\phi_6=\phi_7=\phi_8=1$. With the help of these functions the determining Eqs. (\ref{ch6.p})-(\ref{ch6.q}) can be simplified. The resultant equations for the functions $\psi_1$ turn out to be
\begin{subequations}\label{ch6.compta}
\begin{eqnarray}
g_1\psi_{1x}+g_2\psi_{1y}-\omega_1\psi_1=0,\label{ch6.compta1}\\
h_1\psi_{1x}+h_4\psi_{1y}-\psi_{1xx}=0,\label{ch6.compta2}\\
h_2\psi_{1x}+h_5\psi_{1y}-\psi_{1yy}=0,\label{ch6.compta3}\\
h_3\psi_{1x}+h_6\psi_{1y}-2\psi_{1xy}=0,\label{ch6.compta4}
\end{eqnarray}
\end{subequations}
where $\psi_{1x}\neq0$ and $\psi_{1y}\neq0$. Now, with the help of the first two relations of Eq. (\ref{ch6.compta}) one can equate the values of $\psi_{1xx}$. Then making use of the other relations, we arrive at
\begin{eqnarray}
\psi_{1x}[g_1h_1+g_{1x}+\frac{1}{2}g_2h_3-\omega_1]+\psi_{1y}[g_1h_4+g_{2x}+\frac{1}{2}g_2h_6]=0.\label{ch6.mcomp1}
\end{eqnarray}
Again, equating the value of $\psi_{1yy}$ from first and third relations of Eqs. (\ref{ch6.compta}), we get
\begin{eqnarray}
\psi_{1x}[g_2h_2+g_{1y}+\frac{1}{2}g_1h_3]+\psi_{1y}[g_2h_5+g_{2y}+\frac{1}{2}g_1h_6-\omega_1]=0.\label{ch6.mcomp2}
\end{eqnarray}
For nontrivial solutions for $\psi_{1x}$ and $\psi_{1y}$ to exist, from Eqs. (\ref{ch6.mcomp1}) and (\ref{ch6.mcomp2}) we require that
\begin{eqnarray}
[g_1h_1+g_{1x}+\frac{1}{2}g_2h_3-\omega_1][g_2h_5+g_{2y}+\frac{1}{2}g_1h_6-\omega_1]&=[g_1h_4+g_{2x}+\frac{1}{2}g_2h_6]\nonumber\\
&\times[g_2h_2+g_{1y}+\frac{1}{2}g_1h_3].
\label{ch6.cond1}
\end{eqnarray}

Similarly, the modified determining equations for the function $\psi_2$ are
\begin{subequations}\label{ch6.comptb}
\begin{eqnarray}
g_1\psi_{2x}+g_2\psi_{2y}-\omega_2\psi_2=0,\label{ch6.comptb1}\\
h_1\psi_{2x}+h_4\psi_{2y}-\psi_{2xx}=0,\label{ch6.comptb2}\\
h_2\psi_{2x}+h_5\psi_{2y}-\psi_{2yy}=0,\label{ch6.comptb3}\\
h_3\psi_{2x}+h_6\psi_{2y}-2\psi_{2xy}=0,\label{ch6.comptb4}
\end{eqnarray}
\end{subequations}
where $\psi_{2x}\neq0$ and $\psi_{2y}\neq0$. Following the same procedure for $\psi_2$ as was done for $\psi_1$, we arrive at the relation
\begin{eqnarray}
[g_1h_1+g_{1x}+\frac{1}{2}g_2h_3-\omega_2][g_2h_5+g_{2y}+\frac{1}{2}g_1h_6-\omega_2]&=[g_1h_4+g_{2x}+\frac{1}{2}g_2h_6]\nonumber\\
&\times[g_2h_2+g_{1y}+\frac{1}{2}g_1h_3].
\label{ch6.cond2}
\end{eqnarray}
Equating the right hand sides of Eqs. (\ref{ch6.cond1}) and (\ref{ch6.cond2}) and simplifying, we get
\begin{eqnarray}
g_1h_1+g_{1x}+\frac{1}{2}g_2h_3+g_2h_5+g_{2y}+\frac{1}{2}g_1h_6-\omega_1-\omega_2=0,\label{ch6.cond3}
\end{eqnarray}
provided $\omega_1\neq\omega_2$.

Eq. (\ref{ch6.cond3}) along with Eq. (\ref{ch6.cond1}) or (\ref{ch6.cond2}) can be used to identify the isochronous equations belonging to the general Eq. (\ref{ch6.fact}) by imposing these conditions on the form of the functions $h_i$ and $g_j$. 

%%%%%%%%%%%%%%%%%%%%%%%%%%%%%%%%%%%%%%%%%%%%%%%%%%%%%%%%%%%%%%%%%%%%%%%%%%%%%%%%%%%%%%%%%%%%%%%%%%%%%%%%%%%%%%%%%%%%%%%%%%%%%%%%%%%%%%%%%%%%%%%%%%%%%%%%%%%%%%
\noindent
{\bf Limiting Case:}
It is to be noted that for the scalar quadratic Li\'enard type equation \cite{akt}
\begin{eqnarray}
\ddot{x}+h(x)\dot{x}^2+g(x)=0,\label{sca1}
\end{eqnarray}
where $h(x)$ and $g(x)$ are arbitrary functions of $x$ only, the isochronicity condition from Eqs. (\ref{ch6.cond1}) and (\ref{ch6.cond2}) turns out to be
\begin{eqnarray}
g_x+hg=\omega_1,
\end{eqnarray}
which is exactly the same as has been proved by many authors \cite{musk,chou,saba,chou2,chou3}.

%%%%%%%%%%%%%%%%%%%%%%%%%%%%%%%%%%%%%%%%%%%%%%%%%%%%%%%%%%%%%%%%%%%%%%%%%%%%%%%%%%%%%%%%%%%%%%%%%%%%%%%%%%%%%%%%%%%%%%%%%%%%%%%%%%%%%%%%%%%%%%%%%%%%%%%%%%%%%%
\section{Example of isochronicity}
In this section, we consider a physically interesting example and show that it satisfies the isochronous condition given in the previous section and exhibits amplitude independent periodic solutions. We also find their general solutions.

Let us consider a second order ODE of the form \cite{vkc3}
\begin{eqnarray}
\ddot{x}+\frac{(\dot{x}y-\dot{y}x)^2}{2xy(x-y)}+\omega x=0 ,\quad
\ddot{y}-\frac{(\dot{x}y-\dot{y}x)^2}{2xy(x-y)}+\omega y=0.\label{ch6.isex1}
\end{eqnarray}
Identifying the forms of the functions $h_i$ and $g_j$ as
\begin{eqnarray}
\hspace{-.4cm}h_1=\frac{y}{2x(x-y)},\quad h_2=\frac{x}{2y(x-y)},\quad h_3=-\frac{1}{x-y},\quad g_1=\omega x,\nonumber\\
\hspace{-.4cm}h_4=-\frac{y}{2x(x-y)},\quad h_5=-\frac{x}{2y(x-y)},\quad h_6=\frac{1}{x-y},\quad g_2=\omega y.
\end{eqnarray}
One can check that the above forms of $h_i$ and $g_i$ satisfy the isochronicity condition given by Eqs. (\ref{ch6.cond1}) and (\ref{ch6.cond2}). Hence, Eq. (\ref{ch6.isex1}) can be transformed to the system of coupled simple harmonic oscillator equations with appropriate transformation. The system of coupled simple harmonic oscillator equations can be written as
\begin{eqnarray}
&&[D+\sqrt{-\omega}][D-\sqrt{-\omega}](x+y)=0, \nonumber \\
&&[D+\sqrt{-\omega}][D-\sqrt{-\omega}]\sqrt{xy}=0,
\end{eqnarray} 
where the form of the functions $\psi_{1}$ and $\psi_2$ are obtained by solving Eqs. (\ref{ch6.compta}) and (\ref{ch6.comptb}) consistently. One may check that expanding the above set of equations one can get the system (\ref{ch6.isex1}) under consideration. It means that the transformation $\psi_1=x+y$ and $\psi_2=\sqrt{xy}$ leads the simple harmonic oscillator equation to the desired equation. Hence, the solution of the desired equation can be obtained from the solution of the simple harmonic oscillator equation just by inverting the relation. The solution of simple harmonic oscillator  is 
\begin{eqnarray}
\psi_1=A\sin{(\omega\, t+\delta_1)},\qquad \psi_2=B\sin{(\omega \,t+\delta_2)}.
\end{eqnarray} 
Substituting $\psi_1=x+y$ and $\psi_2=\sqrt{xy}$ in the above equation and solving, we get the solution of Eq. (\ref{ch6.isex1}) as
\begin{eqnarray}
\hspace{-.8cm}x(t)=\frac{1}{2}\left(A\sin{(\omega \,t+\delta_1)}\pm \sqrt{A^2\sin^2{(\omega \,t+\delta_1)}-4B^2\sin^2{(\omega \,t+\delta_2)}} \right),\\
\hspace{-.8cm}y(t)=\frac{1}{2}\left(A\sin{(\omega \,t+\delta_1)}\mp \sqrt{A^2\sin^2{(\omega \,t+\delta_1)}-4B^2\sin^2{(\omega \,t+\delta_2)}} \right).
\end{eqnarray}
where $A,B,\delta_1$ and $\delta_2$ are integration constants.

%%%%%%%%%%%%%%%%%%%%%%%%%%%%%%%%%%%%%%%%%%%%%%%%%%%%%%%%%%%%%%%%%%%%%%%%%%%%%%%%%%%%%%%%%%%%%%%%%%%%%%%%%%%%%%%%%%%%%%%%%%%%%%%%%%%%%%%%%%%%%%%%%%%%%%%%%%%%%%
\section{The case of mixed Li\'enard type equation}
In this section, we consider the addition of a linear velocity term in addition to the quadratic velocity term in Eq. (\ref{ch6.int4}) to get an overview of a more general class of equations.
 
Including the linear velocity term, Eq. (\ref{ch6.int4}) can be written as
\begin{subequations}
\label{mx1}
\begin{eqnarray}
\ddot{x}+h_1(x,y)\dot{x}^2+h_2(x,y)\dot{y}^2+h_3(x,y)\dot{x}\dot{y}+f_1(x,y)\dot{x}+f_2(x,y)\dot{y}+g_1(x,y)=0,\label{mx1.1}  \\
\ddot{y}+h_4(x,y)\dot{x}^2+h_5(x,y)\dot{y}^2+h_6(x,y)\dot{x}\dot{y}+f_3(x,y)\dot{x}+f_4(x,y)\dot{y}+g_2(x,y)=0.\label{mx1.2}
\end{eqnarray}
\end{subequations}
Inclusion of the linear velocity terms will not change the form of ${h_i}^{'}s$ and ${g_j}^{'}s$ and hence the determining equations for $\psi_{1,2},\,\phi_5$ and $\phi_6$ remain unchanged. The only difference is that now the coefficients of $\dot{x}$ and $\dot{y}$ will not be zero but will be given by the functions ${f_l}^{'}s,\,l=1,2,3,4,$. Now, comparing the above equation with (\ref{ch6.fact}) we get the forms of the functions ${f_l}^{'}s,\,l=1,2,3,4,$ as
\begin{subequations}
\label{ch8.fdef}
\begin{align}
{f}_1&=\frac{1}{\delta}[-\phi_6\phi_8 \psi_{2y}(\phi_7(\psi_{1}\phi_{2x}+\psi_{1x}\phi_{2})+\phi_1\phi_5\psi_{1x})+\phi_5 \phi_7\psi_{1y}(\phi_8(\psi_{2}\phi_{4x}+\psi_{2x}\phi_{4})+\phi_3\phi_6\psi_{2x}],\label{ch8.fdefa} \\
{f}_2&=\frac{1}{\delta}[-\phi_6\phi_8 \psi_{2y}(\phi_7(\psi_{1}\phi_{2y}+\psi_{1y}\phi_{2})+\phi_1\phi_5\psi_{1y})+\phi_5 \phi_7\psi_{1y}(\phi_8(\psi_{2}\phi_{4y}+\psi_{2y}\phi_{4})+\phi_3\phi_6\psi_{2y}]\label{ch8.fdefb}, \\
{f}_3&=\frac{1}{\delta}[\phi_6\phi_8 \psi_{2x}(\phi_7(\psi_{1}\phi_{2x}+\psi_{1x}\phi_{2})+\phi_1\phi_5\psi_{1x})-\phi_5 \phi_7\psi_{1x}(\phi_8(\psi_{2}\phi_{4x}+\psi_{2x}\phi_{4})+\phi_3\phi_6\psi_{2x}],\label{ch8.fdef1a}\\
{f}_4&=\frac{1}{\delta}[\phi_6\phi_8 \psi_{2x}(\phi_7(\psi_{1}\phi_{2y}+\psi_{1y}\phi_{2})+\phi_1\phi_5\psi_{1y})-\phi_5 \phi_7\psi_{1x}(\phi_8(\psi_{2}\phi_{4y}+\psi_{2y}\phi_{4})+\phi_3\phi_6\psi_{2y}].\label{ch8.fdef1b}  
\end{align}
\end{subequations}
In the case of Eq. (\ref{ch6.int4}) the left hand sides of Eqs. (\ref{ch8.fdef}) are zero. Hence, we can determine $\phi_2$ and $\phi_4$ easily. To get the form of the functions $\phi_2$ and $\phi_4$ in the present case we proceed in the following manner. 

Multiplying (\ref{ch8.fdefa}) by $\psi_{1x}$ and (\ref{ch8.fdef1a}) by $\psi_{1y}$ and adding, we get
\begin{eqnarray}
\phi_7(\phi_{2x}\psi_1+\phi_2\psi_{1x})=-\phi_5\phi_7(\psi_{1x}f_1+\psi_{1y}f_3)-\phi_1\phi_5\psi_{1x}.\label{ch8.p1a}
\end{eqnarray}
Now, multiplying the above equation by $\psi_1\phi_2$ and using (\ref{ch6.phi13gen2}), we arrive at the following relation
\begin{eqnarray}
\frac{\partial}{\partial x}\left[\frac{\phi_2^2\psi_1^2}{2}\right]=-(\psi_{1x}g_1+\psi_{1y}g_2)\phi_5^2\psi_{1x}-(\psi_{1x}f_1+\psi_{1y}f_3){\phi_2}{\phi_5}\psi_1.\label{ch8.pde1}
\end{eqnarray}
Again, multiplying (\ref{ch8.fdefa}) by $\psi_{2x}$ and (\ref{ch8.fdef1a}) by $\psi_{2y}$ and simplifying in the same manner as we have done above, we find
\begin{eqnarray}
\frac{\partial}{\partial x}\left[\frac{\phi_4^2\psi_2^2}{2}\right]=-(\psi_{2x}g_1+\psi_{2y}g_2)\phi_6^2\psi_{2x}-(\psi_{2x}f_1+\psi_{2y}f_3){\phi_4}{\phi_6}\psi_2.\label{ch8.pde2}
\end{eqnarray}
Proceeding in the same way for Eqs. (\ref{ch8.fdefb}) and (\ref{ch8.fdef1b}), we arrive at the following relations
\begin{eqnarray}
\frac{\partial}{\partial y}\left[\frac{\phi_2^2\psi_1^2}{2}\right]=-(\psi_{1x}g_1+\psi_{1y}g_2)\phi_5^2\psi_{1y}-(\psi_{1x}f_2+\psi_{1y}f_4){\phi_2}{\phi_5}\psi_1,\label{ch8.pde3}\\
\frac{\partial}{\partial x}\left[\frac{\phi_4^2\psi_2^2}{2}\right]=-(\psi_{2x}g_1+\psi_{2y}g_2)\phi_6^2\psi_{2y}-(\psi_{2x}f_2+\psi_{2y}f_4){\phi_4}{\phi_6}\psi_2.\label{ch8.pde4}
\end{eqnarray}
Using the compatibility condition of (\ref{ch8.pde1}) with (\ref{ch8.pde2}) and of (\ref{ch8.pde3}) with (\ref{ch8.pde4}), we can write
\begin{subequations}
\label{ch8.comta}
\begin{align}
\frac{\partial}{\partial y}&\left[(\psi_{1x}g_1+\psi_{1y}g_2)\phi_5^2\psi_{1x}+(\psi_{1x}f_1+\psi_{1y}f_3){\phi_2}{\phi_5}\psi_1\right]\nonumber\\
&\qquad=\frac{\partial}{\partial x}\left[(\psi_{1x}g_1+\psi_{1y}g_2)\phi_5^2\psi_{1y}+(\psi_{1x}f_2+\psi_{1y}f_4){\phi_2}{\phi_5}\psi_1\right],\label{ch8.com2}\\
\frac{\partial}{\partial y}&\left[(\psi_{2x}g_1+\psi_{2y}g_2)\phi_6^2\psi_{2x}+(\psi_{2x}f_1+\psi_{2y}f_3){\phi_4}{\phi_6}\psi_2\right]\nonumber\\
&\qquad=\frac{\partial}{\partial x}\left[(\psi_{2x}g_1+\psi_{2y}g_2)\phi_6^2\psi_{2y}+(\psi_{2x}f_2+\psi_{2y}f_4){\phi_4}{\phi_6}\psi_2\right].\label{ch8.com1}
\end{align}
\end{subequations}
If we define 
\begin{subequations}
\label{ch8.hatgf}
\begin{eqnarray}
&\hat{g_1}=\psi_{1x}g_1+\psi_{1y}g_2, &\hat{g_2}=\psi_{2x}g_1+\psi_{2y}g_2,\label{ch8.hatg}\\
&\hat{f_1}=\psi_{1x}f_1+\psi_{1y}f_3, &\hat{f_2}=\psi_{1x}f_2+\psi_{1y}f_4,\\
&\hat{f_3}=\psi_{2x}f_1+\psi_{2y}f_3, &\hat{f_4}=\psi_{2x}f_2+\psi_{2y}f_4,\label{ch8.hatf}
\end{eqnarray}
\end{subequations}
then Eqs. (\ref{ch8.comta}) can be written as
\begin{eqnarray}
(\hat{f}_2\phi_{2x}- \hat{f}_1\phi_{2y})\phi_5\psi_1&+[(\hat{f}_{2x}-\hat{f}_{1y})\phi_5\psi_1+\hat{f}_2(\phi_{5x}\psi_1+\phi_5\psi_{1x})-\hat{f}_1(\phi_{5y}\psi_1+\phi_5\psi_{1y})]\phi_2\nonumber\\
&+{\phi_5}^2(\hat{g}_{1x}\psi_{1y}-\hat{g}_{1y}\psi_{1x})+2\hat{g}_{1}\phi_5(\phi_{5x}\psi_{1y}-\phi_{5y}\psi_{1x})=0,\label{ch8.phi2a}\\
(\hat{f}_4\phi_{4x}- \hat{f}_3\phi_{2y})\phi_6\psi_2&+[(\hat{f}_{4x}-\hat{f}_{3y})\phi_6\psi_2+\hat{f}_4(\phi_{6x}\psi_2+\phi_6\psi_{2x})-\hat{f}_3(\phi_{6y}\psi_2+\phi_6\psi_{2y})]\phi_4\nonumber\\
&+{\phi_6}^2(\hat{g}_{2x}\psi_{2y}-\hat{g}_{2y}\psi_{2x})+2\hat{g}_{2}\phi_6(\phi_{6x}\psi_{2y}-\phi_{6y}\psi_{2x})=0\label{ch8.phi4}.
\end{eqnarray}
The above set equations are the determining equations for $\phi_2$ and $\phi_4$. As mentioned, the determining equations for $\psi_{1,2}$ and $\phi_5$ and $\phi_6$ are the same as in the case of Eq. (\ref{ch6.int4}) and for $\psi_{1,2}$ it is given by (\ref{ch6.psi}), and $\phi_5$ and $\phi_6$ are given by Eqs. (\ref{ch6.phi1})-(\ref{ch6.phi2}). Hence, proceeding in the same way as discussed in Secs. 4 and 5 one can determine the forms of the function $\psi_{1,2},\,\phi_5$ and $\phi_6$. Substituting $\psi_{1,2},\,\phi_5$ and $\phi_6$ in Eqs. (\ref{ch8.comta}) and solving the resultant PDEs the functions $\phi_2$ and $\phi_4$ can be obtained. With the help of the known forms in Eq. (\ref{ch6.phi13gen2}) the forms of $\phi_1$ and $\phi_3$ can be fixed in terms of $\phi_7$ and $\phi_8$.  

%%%%%%%%%%%%%%%%%%%%%%%%%%%%%%%%%%%%%%%%%%%%%%%%%%%%%%%%%%%%%%%%%%%%%%%%%%%%%%%%%%%%%%%%%%%%%%%%%%%%%%%%%%%%%%%%%%%%%%%%%%%%%%%%%%%%%%%%%%%%%%%%%%%%%%%%%%%%%%
\section{Conclusion}
In this paper, we have developed a systematic and self contained procedure which enables us to analyse the factorization of a rather general class of coupled quadratic Li\'enard type equations in terms of first order differential operators. In this way, we have shown that the factorized form for the given equation can be obtained in a systematic and simple way. This reduces the problem of finding the solution of the equations belonging to this class to the problem of solving a set of first order differential equations. To demonstrate the effectiveness of this procedure we considered coupled ML oscillator equation and factorized it systematically. In addition to this, we have also considered the isochronous properties of this equation and deduced the isochronicity condition for it. With the help of this condition one can identify the forms of Eq. (\ref{ch6.fact}) exhibiting isochronous property. An example of physical interest is also discussed. Finally, we have also extended the procedure to the case of coupled mixed type of Li\'enard equations by including linear velocity terms in addition to the quadratic velocity terms.    

%%%%%%%%%%%%%%%%%%%%%%%%%%%%%%%%%%%%%%%%%%%%%%%%%%%%%%%%%%%%%%%%%%%%%%%%%%%%%%%%%%%%%%%%%%%%%%%%%%%%%%%%%%%%%%%%%%%%%%%%%%%%%%%%%%%%%%%%%%%%%%%%%%%%%%%%%%%%%%
\section{Acknowledgments}
AKT and SNP are grateful to the Centre for Nonlinear Dynamics, Bharathidasan University, Tiruchirappalli, for warm hospitality. The work of SNP forms part of a Department of Science and Technology, Government of India, sponsored research project. The work of VKC and ML is supported by a DST-IRHPA research project. The work of ML is also supported by DST-Ramanna Fellowship program and a DAE Raja Ramanna Fellowship.

%%%%%%%%%%%%%%%%%%%%%%%%%%%%%%%%%%%%%%%%%%%%%%%%%%%%%%%%%%%%%%%%%%%%%%%%%%%%%%%%%%%%%%%%%%%%%%%%%%%%%%%%%%%%%%%%%%%%%%%%%%%%%%%%%%%%%%%%%%%%%%%%%%%%%%%%%%%%%%
\appendix
\begin{appendix}
%%%%%%%%%%%%%%%%%%%%%%%%%%%%%%%%%%%%%%%%%%%%%%%%%%%%%%%%%%%%%%%%%%%%%%%%%%%%%%%%%%%%%%%%%%%%%%%%%%%%%%%%%%%%%%%%%%%%%%%%%%%%%%%%%%%%%%%%%%%%%%%%%%%%%%%%%%%%%%
\section{Factorization of scalar case corresponding to Eq. (\ref{ch6.int4})}
In this appendix, we consider the scalar case corresponding to Eq. (\ref{ch6.int4}), that is 
\begin{eqnarray}
\ddot{x}+h(x)\dot{x}^2+g(x)=0,\label{ap1}
\end{eqnarray}
where $h(x)$ and $g(x)$ are functions of $x$, and discuss how to get the factorized form for Eq. (\ref{ap1}). We will also consider a specific example  belonging to Eq. (\ref{ap1}).

%%%%%%%%%%%%%%%%%%%%%%%%%%%%%%%%%%%%%%%%%%%%%%%%%%%%%%%%%%%%%%%%%%%%%%%%%%%%%%%%%%%%%%%%%%%%%%%%%%%%%%%%%%%%%%%%%%%%%%%%%%%%%%%%%%%%%%%%%%%%%%%%%%%%%%%%%%%%%%
\subsection{Factorization of Eq. (\ref{ap1})}
To start with, let us assume that Eq. (\ref{ap1}) can be factorized in the form
\begin{eqnarray}
[\phi_4(x)D-\phi_3(x)][\phi_2(x)D-\phi_1(x)]\psi(x)=0,\label{ap2}
\end{eqnarray}
where ${\phi_k}^{'}s,\,k=1,2,3,4,$ and $\psi(x)$ are unknown functions of $x$ to be determined. It is to be noted that one can always absorb $\phi_4$ in $\phi_3$ by redefining the functions. Hence, we define $\tilde{\phi}_3=\frac{\phi_3}{\phi_4}$. Under this definition Eq. (\ref{ap2}) can be written as
\begin{eqnarray}
[D-\tilde{\phi}_3(x)][\phi_2(x)D-\phi_1(x)]\psi(x)=0.\label{ap3}
\end{eqnarray}
Now, expanding Eq. (\ref{ap3}) and comparing the latter with the coefficients of various powers of $\dot{x}$ of Eq. (\ref{ap1}), we get
\begin{eqnarray}
(\phi_2\psi_x)_x-h\,\phi_2\psi_x=0,\label{ap4.1}\\
\phi_1\tilde{\phi}_3\psi-\phi_2\psi_x\,g=0,\label{ap4.2}
\end{eqnarray} 
and 
\begin{eqnarray}
(\phi_1\psi)_x+\phi_2\tilde{\phi}_3\psi_x=0.\label{ap4.3}
\end{eqnarray}
In order to get the factorized form we need to identify the unknown functions, that is $\phi_1,\phi_2,\tilde{\phi}_3$ and $\psi$. To deduce suitable forms for these unknowns we need to solve Eqs. (\ref{ap4.1})-(\ref{ap4.3}) consistently. 

Solving Eq. (\ref{ap4.1}), we get
\begin{eqnarray}
\phi_2\psi_x=c_1e^{\int{h\,dx}},\label{ap5}
\end{eqnarray}
where $c_1$ is an arbitrary constant. Using Eq. (\ref{ap5}) in Eq. (\ref{ap4.2}), we find
\begin{eqnarray}
\phi_1\tilde{\phi}_3\psi=g\,c_1\,e^{\int{h\,dx}}.\label{ap6}
\end{eqnarray}
Similarly, using (\ref{ap5}) in (\ref{ap4.3}), we get
\begin{eqnarray}
\tilde{\phi}_3\,e^{\int{h\,dx}}=-(\phi_1\psi)_x.\label{ap7}
\end{eqnarray}
Using Eq. (\ref{ap6}) in (\ref{ap7}), we get
\begin{eqnarray}
\phi_1\psi(\phi_1\psi)_x=-g\,c_1^2e^{2\int{h\,dx}}.\label{ap8}
\end{eqnarray}
Solving above equation, we get
\begin{eqnarray}
\phi_1\psi=\sqrt{c_2-2c_1^2\int{g\,e^{2\int{h\,dx}}dx}},\label{ap9}
\end{eqnarray}
where $c_2$ is a constant. Hence, Eqs. (\ref{ap5}), (\ref{ap7}) and (\ref{ap9}) can be used to identify the suitable form of the unknown functions. Now, it is clear from Eqs. (\ref{ap3}) and (\ref{ap9}) that we have freedom to choose the form of the function $\phi_1$ in order to obtain the factorized form (\ref{ap3}). Hence, we consider $\phi_1$ as a function of $x$, say $M(x)$. Then from Eq. (\ref{ap9}), the form of $\psi$ can be written as
\begin{eqnarray}
\psi =\frac{\sqrt{c_2-2c_1^2\int g\,e^{2\int h \, dx}dx}}{M}.\label{ap101}
\end{eqnarray}
With the above form of $\psi$, the form of $\phi_2$ can be written from Eq. (\ref{ap5}) as
\begin{eqnarray}
\phi_2=-\frac{c_1\,M^2\,e^{\int h\,dx}\sqrt{c_2-2c_1^2\int e^{2\int h\,dx}g\,dx}}{g\,M\,c_1^2\,e^{2\int h\,dx}+\left(c_2-2c_1^2\int e^{2\int h\,dx}g\,dx\right) M'},\label{ap102}
\end{eqnarray} 
where $'$ denotes differentiation with respect to $x$. Using (\ref{ap101}) in (\ref{ap6}) the form of $\tilde{\phi}_3$ turn out to be
\begin{eqnarray}
\tilde{\phi}_3=\frac{c_1\,g\,e^{\int h\,dx}}{\sqrt{c_2-2c_1^2\int ge^{2\int h\,dx}dx}}.\label{ap103}
\end{eqnarray}
These above form of the functions $\psi,\,\phi_2$ and $\tilde{\phi}_3$ given by Eqs. (\ref{ap101}), (\ref{ap102}) and (\ref{ap103}) and $\phi_1=M$ satisfies Eq. (\ref{ap1}). It means that even for the arbitrary form of $\phi_1$ the above forms of the functions $\psi,\,\phi_2$ and $\tilde{\phi}_3$ give the suitable factorized form for Eq. (\ref{ap1}). Hence, we can consider $\phi_1$ as a constant. To illustrate this procedure, in the following, we consider an example of physical and mathematical interest. 

%%%%%%%%%%%%%%%%%%%%%%%%%%%%%%%%%%%%%%%%%%%%%%%%%%%%%%%%%%%%%%%%%%%%%%%%%%%%%%%%%%%%%%%%%%%%%%%%%%%%%%%%%%%%%%%%%%%%%%%%%%%%%%%%%%%%%%%%%%%%%%%%%%%%%%%%%%%%%%
As an example, we consider the ML oscillator equation \cite{cari,cari1}
\begin{eqnarray}
\ddot{x}-\frac{\lambda x}{1+\lambda x^2}\,\dot{x}^2+\frac{\omega x}{1+\lambda x^2}=0,\label{ap12}
\end{eqnarray}
where $\lambda$ and $\omega$ are arbitrary parameters. Comparing above equation with Eq. (\ref{ap1}), we get
\begin{eqnarray}
h=-\frac{\lambda x}{1+\lambda x^2},\qquad g=\frac{\omega x}{1+\lambda x^2}.\label{ap13}
\end{eqnarray}
If, Eq. (\ref{ap13}) can be written in factorized form as Eq. (\ref{ap3}), then we can write the determining equations for the unknowns with the help of Eqs. (\ref{ap4.1})-(\ref{ap4.3}). To get the factorized form we consider $\phi_1$ as a constant, say $\sqrt{-\omega }$. Then solving Eq. (\ref{ap9}) for $\psi$, we get
\begin{eqnarray}
\psi=\frac{x}{\sqrt{1+\lambda x^2}},
\end{eqnarray} 
where $c_1=1$ and $c_2=-\frac{\omega}{\lambda}$. Now, $\phi_2$ can be obtained by solving Eq. (\ref{ap5}) as
\begin{eqnarray}
\phi_2=1+\lambda x^2
\end{eqnarray}
and $\tilde{\phi}_3$ can be obtained from (\ref{ap6}) as
\begin{eqnarray}
\tilde{\phi}_3=-\frac{\sqrt{-\omega}}{1+\lambda x^2}.
\end{eqnarray}
With the help of the obtained forms of the unknown functions the factorized form for ML oscillator can be written as
\begin{eqnarray}
\left[(1+\lambda x^2)D\pm{\sqrt{-\omega }}\right]\left[(1+\lambda x^2)D\mp\sqrt{-\omega }\right]\,\frac{x}{\sqrt{1+\lambda x^2}}=0,\label{ap15}
\end{eqnarray}
where we have used the relation $\tilde{\phi}_3=\frac{\phi_3}{\phi_4}$ and $\phi_4=1+\lambda x^2$.

To prove the integrability of Eq. (\ref{ap12}), we proceed in the same way as was done in Sec. 6.2. Hence, we can write Eq. (\ref{ap15}) as
\begin{subequations}
\label{ap16}
\begin{eqnarray}
&&\left[\tilde{D}+\sqrt{-\omega}\right]\left[\tilde{D}-\sqrt{-\omega}\right]\,\frac{x}{\sqrt{1+\lambda x^2}}=0,\label{ap16.1}\\
&&\left[\tilde{D}-\sqrt{-\omega}\right]\left[\tilde{D}+\sqrt{-\omega}\right]\,\frac{x}{\sqrt{1+\lambda x^2}}=0,\label{ap16.2}
\end{eqnarray}
\end{subequations}
where $\tilde{D}=(1+\lambda x^2)D$. The above equation can be written as
\begin{eqnarray}
\hspace{-.9cm}\frac{\tilde{D}\left[\tilde{D}[\frac{x}{\sqrt{1+\lambda x^2}}]-\frac{\sqrt{-\omega}\,x}{\sqrt{1+\lambda x^2}}\right]}
{\tilde{D}[\frac{x}{\sqrt{1+\lambda x^2}}]-\frac{\sqrt{-\omega}\,x}{\sqrt{1+\lambda x^2}}}
+\frac{\tilde{D}\left[\tilde{D}[\frac{x}{\sqrt{1+\lambda x^2}}]+\frac{\sqrt{-\omega}\,x}{\sqrt{1+\lambda x^2}}\right]}
{\tilde{D}[\frac{x}{\sqrt{1+\lambda x^2}}]+\frac{\sqrt{-\omega}\,x}{\sqrt{1+\lambda x^2}}}=0.\label{ap17}
\end{eqnarray}
Integrating Eq. (\ref{ap17}), we get
\begin{eqnarray}
I_1=\frac{\dot{x}^2+\omega x^2}{1+\lambda x^2},
\end{eqnarray}
where $I_1$ is the first integral for the scalar ML oscillator equation \cite{vkc4}.

%%%%%%%%%%%%%%%%%%%%%%%%%%%%%%%%%%%%%%%%%%%%%%%%%%%%%%%%%%%%%%%%%%%%%%%%%%%%%%%%%%%%%%%%%%%%%%%%%%%%%%%%%%%%%%%%%%%%%%%%%%%%%%%%%%%%%%%%%%%%%%%%%%%%%%%%%%%%%%
\section{Factorization of scalar case corresponding to Eq. (\ref{int1})}
Now, we consider the scalar mixed \cite{akt2,nucci2} case by including an $\dot{x}$ term to Eq. (\ref{ap1}), that is 
\begin{eqnarray}
\ddot{x}+h(x)\dot{x}^2+f(x)\dot x+g(x)=0,\label{man1}
\end{eqnarray}
where $f(x),\,g(x)$ and $h(x)$ are arbitrary functions of $x$, and discuss how to get the factorized form for Eq. (\ref{man1}). For this purpose, we follow the same procedure used in Appendix A earlier. To start with, we assume that Eq. (\ref{man1}) can be factorized in the form given by Eq. (\ref{ap3}). Now, expanding it and comparing the resultant equation with the coefficients of various powers of $\dot{x}$ of Eq. (\ref{man1}), we get
\begin{eqnarray}
(\phi_2\psi_x)_x-h\,\phi_2\psi_x=0,\label{man4.1}\\
\phi_1\tilde{\phi}_3\psi-\phi_2\psi_x\,g=0,\label{man4.2}
\end{eqnarray} 
and 
\begin{eqnarray}
(\phi_1\psi)_x+\phi_2\tilde{\phi}_3\psi_x+f\phi_2\psi_x=0.\label{man4.3}
\end{eqnarray}
It is to be noted that Eqs. (\ref{man4.1}) and (\ref{man4.2}) are exactly the same as Eqs. (\ref{ap4.1}) and (\ref{ap4.2}), respectively, while Eq. (\ref{man4.3}) differs from Eq. (\ref{ap4.3}). Hence, we can deduce the forms of $\phi_2$ and $\tilde{\phi}_3$ to be the same as given by Eq. (\ref{ap5}) and (\ref{ap6}), provided $\phi_1$ and $\psi$ are known. The form of the product function $(\phi_1\psi)$ can be fixed by solving Eq. (\ref{man4.3}). Now, using Eqs. (\ref{ap5}) and (\ref{ap6}) in Eq. (\ref{man4.3}), we get
\begin{eqnarray}
(\phi_1\psi)_x+\frac{g\,c_1^2e^{2\int{h\,dx}}}{\phi_1\psi}+f\,c_1e^{\int{h\,dx}}=0.\label{man8}
\end{eqnarray}
Eq. (\ref{man8}) is of the form of the Abel equation of the second kind, that is
\begin{eqnarray}
\xi \xi'+F(x)\xi+G(x)=0,\quad '\equiv \frac{d}{dx}.\label{man11}
\end{eqnarray}
A general condition for the seperability \cite{ico} of Eq. (\ref{man11}) is known as
\begin{eqnarray}
\left(\frac{G}{F}\right)'=\delta F,\label{man11.1}
\end{eqnarray}
where $\delta$ is an arbitrary constant. Under this condition and with the change of dependent variable $\xi=\left(\frac{G}{F}\right)\frac{1}{w}$, we get the separable equation in the form
\begin{eqnarray}
w'=\frac{F^2}{G}w(w^2+w+\delta),
\end{eqnarray}
which is integrable. Hence, we conclude that equations belonging to Eq. (\ref{man1}) can be factorized into the form (\ref{ap2}) if Eq. (\ref{man8}) can be solved.

%%%%%%%%%%%%%%%%%%%%%%%%%%%%%%%%%%%%%%%%%%%%%%%%%%%%%%%%%%%%%%%%%%%%%%%%%%%%%%%%%%%%%%%%%%%%%%%%%%%%%%%%%%%%%%%%%%%%%%%%%%%%%%%%%%%%%%%%%%%%%%%%%%%%%%%%%%%%%%
As an example to this class of equations, we consider the damped ML oscillator \cite{glad123}, that is
\begin{eqnarray}
\ddot{x}-\frac{\lambda x}{1+\lambda x^2}\,\dot{x}^2+\frac{\alpha}{1+\lambda x^2}\dot{x}+\frac{\lambda_1 x}{1+\lambda x^2}=0,\label{man12}
\end{eqnarray}
where $\lambda,\,\alpha$ and $\lambda_1$ are arbitrary parameters. Comparing Eqs. (\ref{man12}) and (\ref{man1}), we get
\begin{eqnarray}
h=-\frac{\lambda x}{1+\lambda x^2},\qquad f=\frac{\alpha}{1+\lambda x^2},\quad g=\frac{\lambda_1 x}{1+\lambda x^2},\label{man13}
\end{eqnarray}
which indeed satisfy the separability condition (\ref{man11.1}). Now, rewriting Eq. (\ref{man8}) for this case, we get
\begin{eqnarray}
(\phi_1\psi)(\phi_1\psi)_x+\frac{c_1\alpha}{(1+\lambda x^2)^{\frac{3}{2}}}(\phi_1\psi)+\frac{c_1^2\lambda_1x}{(1+\lambda x^2)^2}=0.\label{man14}
\end{eqnarray}
A particular solution to the above equation can be written as
\begin{eqnarray}
\phi_1\psi=\frac{a\sqrt{\lambda_1}x}{\sqrt{1+\lambda x^2}},\label{man114}
\end{eqnarray}
where we have considered $c_1=1$ and $a$ is defined by the relation $\alpha=-\frac{\sqrt{\lambda_1}(a^2+1)}{a}$. Now, we consider $\psi$ as $\frac{x}{1+\lambda x^2}$, then $\phi_1$ can be fixed as $\phi_1=a\sqrt{\lambda_1}$. The remaining functions, that is $\phi_2$ and $\tilde{\phi}_3$ can be obtained from Eqs. (\ref{ap5}) and (\ref{ap6}) as
\begin{eqnarray}
\phi_2=1+\lambda x^2,\label{man16}
\end{eqnarray}
and 
\begin{eqnarray}
{\phi}_3=\frac{\sqrt{\lambda_1}}{a(1+\lambda x^2)}.\label{man17}
\end{eqnarray}
Substituting the obtained forms of the functions $\phi_1,\,\phi_2,\,\tilde{\phi}_3$ and $\psi$ in Eq. (\ref{ap3}) and rewriting the resultant expression, we get the factorized form for Eq. (\ref{man1}) as
\begin{eqnarray}
\left[(1+\lambda x^2)D-{\frac{\sqrt{\lambda_1}}{a}}\right]\left[(1+\lambda x^2)D- a\sqrt{\lambda_1}\right]\,\frac{x}{\sqrt{1+\lambda x^2}}=0.\label{man18}
\end{eqnarray}
Similarly, choosing another particular solution $\phi_1\psi=\frac{\sqrt{\lambda_1}x}{a\sqrt{1+\lambda x^2}}$ to Eq. (\ref{man14}), we can write Eq. (\ref{man12}) as
\begin{eqnarray}
\left[(1+\lambda x^2)D-\sqrt{\lambda_1}a\right]\left[(1+\lambda x^2)D- {\frac{\sqrt{\lambda_1}}{a}}\right]\,\frac{x}{\sqrt{1+\lambda x^2}}=0.\label{man18.1}
\end{eqnarray}

To prove the integrability of the damped ML oscillator (\ref{man12}) we follow the same procedure as that of the scalar ML oscillator case and we obtain 
\begin{eqnarray}
\frac{a}{\sqrt{\lambda_1}}\frac{\tilde{D}\left[\tilde{D}[\frac{x}{\sqrt{1+\lambda x^2}}]-a\sqrt{\lambda_1}\frac{x}{\sqrt{1+\lambda x^2}}\right]}
{\tilde{D}[\frac{x}{\sqrt{1+\lambda x^2}}]-a\sqrt{\lambda_1}\frac{x}{\sqrt{1+\lambda x^2}}}
-\frac{1}{a\sqrt{\lambda_1}}\frac{\tilde{D}\left[\tilde{D}[\frac{x}{\sqrt{1+\lambda x^2}}]-\frac{\sqrt{\lambda_1}}{a}\frac{x}{\sqrt{1+\lambda x^2}}\right]}
{\tilde{D}[\frac{x}{\sqrt{1+\lambda x^2}}]-\frac{\sqrt{\lambda_1}}{a}\frac{x}{\sqrt{1+\lambda x^2}}}=0,\label{man19}
\end{eqnarray}
where $\tilde{D}=(1+\lambda x^2)D$. Integrating Eq. (\ref{man19}), we get
\begin{eqnarray}
I_1=\left[\frac{\dot{x}-a\sqrt{\lambda_1}x}{\sqrt{1+\lambda x^2}}\right]^a\left[\frac{a\dot{x}-\sqrt{\lambda_1}x}{a\sqrt{1+\lambda x^2}}\right]^{-\frac{1}{a}},
\end{eqnarray}
where $I_1$ is the first integral for the scalar damped ML oscillator equation as shown in Ref. \cite{glad123}.
%%%%%%%%%%%%%%%%%%%%%%%%%%%%%%%%%%%%%%%%%%%%%%%%%%%%%%%%%%%%%%%%%%%%%%%%%%%%%%%%%%%%%%%%%%%%%%%%%%%%%%%%%%%%%%%%%%%%%%%%%%%%%%%%%%%%%%%%%%%%%%%%%%%%%%%%%%%%%%
\end{appendix}

%%%%%%%%%%%%%%%%%%%%%%%%%%%%%%%%%%%%%%%%%%%%%%%%%%%%%%%%%%%%%%%%%%%%%%%%%%%%%%%%%%%%%%%%%%%%%%%%%%%%%%%%%%%%%%%%%%%%%%%%%%%%%%%%%%%%%%%%%%%%%%%%%%%%%%%%%%%%%%

\end{document}